# Reactive Glass–Metal Interaction under Ambient Conditions Enables Surface Modification of Gold Nano-islands

Sinorul Haque,[a,b#] Shweta R. Keshri,[c,#] G. Ganesh,[d] Kaustuv Chatterjee,[a,b] Shubhangi Majumdar,[e] Sudheer Ganisetti,[f] Indrajeet Mandal,[g] Dudekula Althaf Basha,[h] Prabir Pal,[a,b] Pramit K Chowdhury,[e] Niharika Joshi,[i] Subrahmanyam Sappati,[j] Nitya Nand Gosvami,[c] Eswaraiah Varrla,[d*] N. M. Anoop Krishnan,[f*] Amarnath R. Allu,[a,b*]

[a] CSIR-Central Glass and Ceramic Research Institute, Kolkata, 700032, India
[b] Academy of Scientific and Innovative Research (AcSIR), Ghaziabad 201002, India
[c] Department of Materials Science and Engineering, Indian Institute of Technology Delhi, Hauz Khas, New Delhi 110016, India
[d] Sustainable Nanomaterials and Technologies Lab, Department of Physics and Nanotechnology, SRM Institute of Science and Technology, Kattankulathur, Chengalpattu, Tamil Nadu 603203, India
[e] Department of Chemistry, Indian Institute of Technology Delhi, Hauz Khas, New Delhi 110016, India
[f] Department of Civil Engineering, Indian Institute of Technology Delhi, New Delhi 110016, India
[g] School of Interdisciplinary Research, Indian Institute of Technology Delhi, Hauz Khas, New Delhi 110016, India
[h] Department of Metallurgy Engineering and Materials Science, Indian Institute of Technology Indore, Simrol, Indore, 453552, India
[i] Catalysis and Inorganic Chemistry Division, CSIR National Chemical Laboratory, Pune-411008, India
[j] BioTechMed Center, and Department of Pharmaceutical Technology and Biochemistry, ul. Narutowicza 11/12 Gdańsk University of Technology, 80-233 Gdańsk, Poland

\# These authors contributed equally to this work

**Corresponding author:**

E-mail:
eswarail@srmist.edu.in **(Eswaraiah Varrla)**
krishnan@iitd.ac.in **(N.M.A. Krishnan)**
alluamarnathreddy@gmail.com; aareddy@cgcri.res.in (**A. R. Allu**)
Fax: +91-33-24730957; Tel: +91-33-23223370

**Abstract:**

Stabilizing gold nanoparticles with tunable surface composition via reactive metal–support interactions under ambient conditions remains a significant challenge. We discovered that a reactive glass–metal interaction (RGMI) at 550 °C under ambient conditions—driven by the intrinsic catalytic activity of gold nano-islands (GNIs) and the unique properties of sodium aluminophosphosilicate glass, including its chemical composition, molar volume, and high $Na^+$ ion mobility—enables the formation of robustly anchored GNIs with altered surface compositions. Comprehensive characterization reveals that the adsorption of Na and P at the GNI surfaces induces lattice distortions in the Au(111) planes. Additionally, a smooth GNI–glass interface significantly influences the hot carrier dynamics of the GNIs. Altogether, RGMI presents a versatile strategy for engineering stable, multi-element nanostructures with potential applications in heterogeneous catalysis, sensing, and optoelectronics.

Owing to their distinctive nanoscale dimensions, metal nanoparticles (MNPs) exhibit exceptional surface-to-volume ratios and quantum-confined electronic properties that drive their remarkable performance in catalysis, plasmonic sensing, and biomedical imaging (*1–5*). Despite these advantages, thermodynamic instability remains a fundamental challenge (*6*), as nanoparticle coalescence and Ostwald ripening progressively degrade catalytic activity. Conventional stabilization strategies—including immobilization on solid supports, embedding in one-dimensional tubular or three-dimensional porous frameworks, and encapsulation within passivating films (*7–9*)—inevitably compromise active site accessibility while introducing synthetic complexities and costs that limit industrial viability. This has prompted intensive research toward developing innovative support systems that not only prevent nanoparticle sintering but also enhance or introduce intrinsic functionality and catalytic efficiency (*6, 10, 11*).

Support materials notably influence nanoparticle properties through strong metal-support interactions (SMSI), modifying structural morphology, electronic band structure, and interfacial chemistry (*2, 12–17*). A particularly intriguing development is the emergence of reactive metal-support interactions (RMSI), wherein support materials facilitate the formation of multielement nanoparticles (MENPs) through chemical exchange at the metal-support interface (*18, 19*). This phenomenon enables the synthesis of unique multi/bimetallic architectures otherwise unattainable through conventional methods, producing novel electronic and geometric configurations at catalytically active sites (*17, 19, 20*). The resulting alterations in surface composition and atomic arrangement significantly enhance molecular adsorption-desorption kinetics, critical for catalytic and sensing applications. However, current RMSI-based approaches present considerable challenges: they typically require reducible oxide or non-oxide supports under hydrogen-rich environments, demand extreme processing temperatures (>800 °C) that compromise nanoparticle dispersion, and involve resource-intensive procedures characterized by costly precursors, intricate protocols, and extended synthesis durations (*17, 20–23*).

Here, we demonstrate that engineered non-reducible sodium aluminophosphosilicate (NAPS) glass substrates enable synthesis of ultra-stable gold nano-islands (GNIs) through thermal dewetting of direct current (DC)-sputtered gold films (see Schematic S1). Remarkably, at moderate temperatures (~550 °C) under ambient conditions, interfacial reactions between NAPS glass and GNIs drive glass element migration onto nanoparticle surfaces, generating multi-element gold nano-islands (MEGNIs). Our systematic investigation across compositionally varied glass substrates reveals that beyond the catalytic properties of GNIs, the chemical composition, thermal behavior and physical characteristics of the glass matrix critically determine interface dynamics. This unprecedented transformation—particularly striking given gold's inherent chemical inertness due to its low work function and surface energy (*9*), coupled with the traditionally non-reactive nature of glass—establishes a novel pathway for nanoparticle surface engineering. The significance extends beyond fundamental materials science, as MENPs exhibit enhanced catalytic efficiency and plasmonic properties compared to their monometallic counterparts (*24–30*), opening opportunities for advanced MENP-based technologies in heterogeneous catalysis, sensing, and optoelectronic applications.

*Characterization of Reactive Metal-Glass Support Interactions*

We engineered a series of sodium aluminophosphosilicate (NAPS-X) glasses with systematically varied silica content (X = 5, 15, 25 and 30 mol%), maintaining fixed $Na_2O$ concentration while adjusting $Al_2O_3$ and $P_2O_5$ ratios (detailed compositions in Supporting

Information S2). These substrates underwent gold film deposition via DC sputtering, followed by thermal treatment at 550 °C for 15 minutes in ambient atmosphere—a notably simple thermal dewetting process that transforms continuous films into embedded GNIs (Schematic S1). The resulting composite materials, designated NAPS-X-Au (Figure 1A), serve as platforms for investigating glass-metal interfacial chemistry.

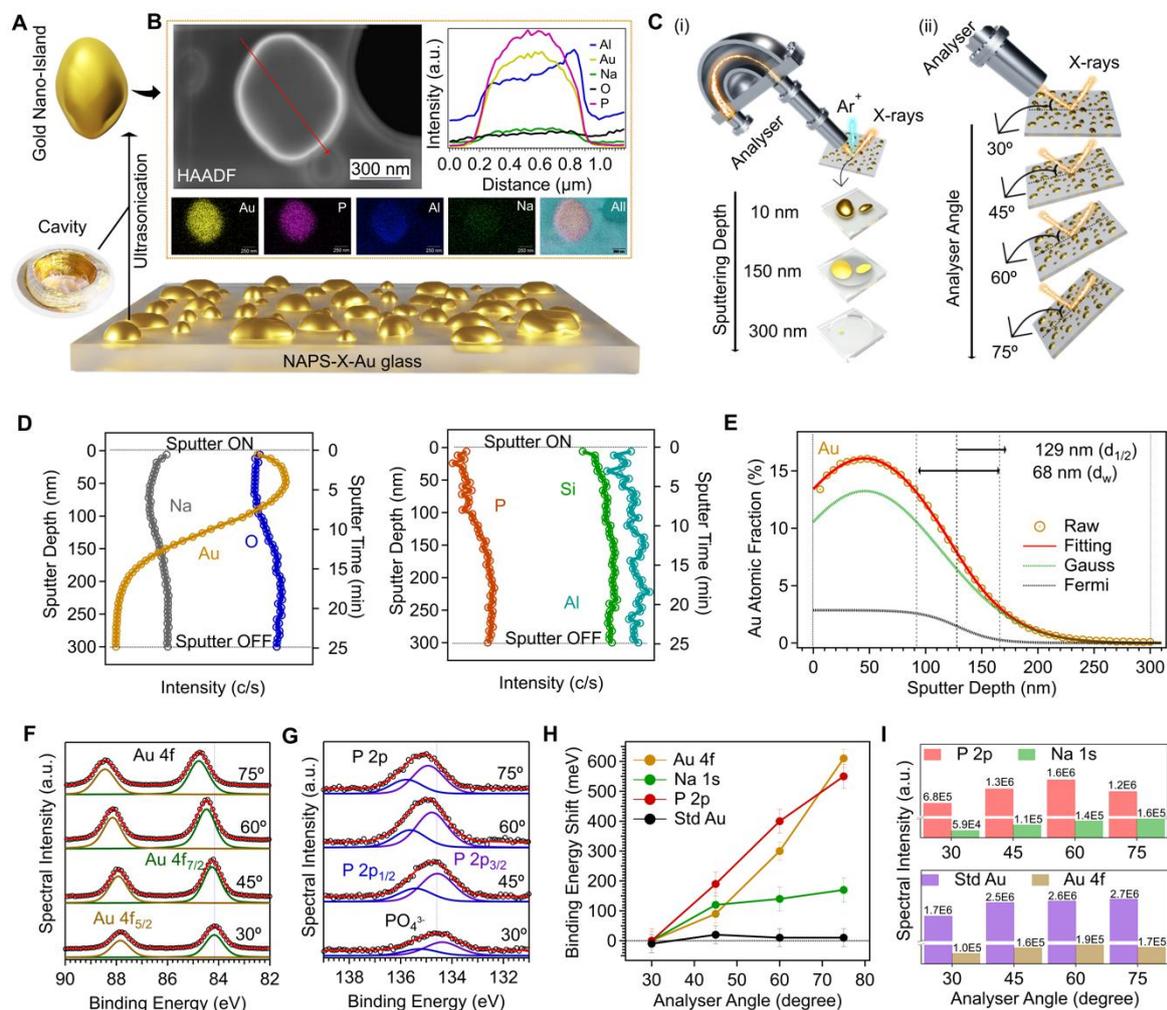

**Figure 1: Reactive glass-metal interaction (RGMI).** *(A) Schematic picture demonstrating the firmly embedded gold nano-islands (GNIs) on the surfaces of NAPS-X-Au glasses. Implementation of high power ultrasonication results the extraction of GNI, which obviously leaves the cavity on the surface of the NAPS-X-Au glass. (B) HAADF-STEM images, along with elemental line profiles and mappings, for GNI extracted from NAPS-15-Au. (C) Schematic picture demonstrating the (i) XPS depth profiling and (ii) angle dependent XPS measurements. (D) Variations in the integrated area under the XPS high-resolution peaks for Au, Na, O, P, Si, and Al as a function of sputtering depth for NAPS-15-Au. (E) Changes in the atomic fraction of Au with sputtering depth, fitted using a convolution of Fermi and Gauss functions for NAPS-15-Au. $d_{1/2}$ is the sputter depth at half the signal intensity maxima, and $d_w$ is the width of the interface between the glass and GNIs. High resolution (F) Au 4f and (G) P 2p XPS spectra at various analyser angle. (H) Angle-dependent shift in peak positions of Au from Standard Au film, and, Au, P, and Na from NAPS-15-Au. (I) Spectral Intensity of Au 4f, Na 1s, and P2p as a function of probe depth on increase of analyser angle compared against Au 4f of standard Au foil.*

Field Emission Scanning Electron Microscopy (FESEM) (Figures S1) and three-dimensional Atomic Force Microscopy (AFM-3D) (Figures S2) confirm thermal dewetting transforms continuous gold films into discrete islands with irregular morphology across all substrates. X-ray diffraction (XRD) establishes the face-centered cubic (FCC) crystalline structure of these GNIs, while UV-VIS spectroscopy reveals characteristic localized surface plasmon resonance (LSPR) behavior (Section S5, Figure S3). Remarkably, Energy-Dispersive X-ray Spectroscopy (EDS) line-scans detect significant phosphorus enrichment within individual gold islands (Figure S1). To elucidate their chemical composition, we extracted GNIs from NAPS-15-Au and NAPS-30-Au substrates via ultrasonication (see methods in supporting) for detailed electron microscopy analysis. Scanning Transmission Electron Microscopy (STEM)-in-SEM imaging, high-angle annular dark-field (HAADF) imaging, EDS elemental line profiles, and EDS elemental mapping shown in Figures 1B and S4 for NAPS-15-Au and NAPS-30-Au, respectively, unequivocally demonstrates surface enrichment with phosphorus, aluminum, and sodium. The homogeneous distribution of these elements without phase segregation indicates systematic migration from the glass substrate to the nanoisland surface. Notably, all islands consistently incorporate phosphorus regardless of substrate composition—a characteristic feature typically observed in RMSI (*31, 32*). Given our use of non-reducible glass as an active support material, we designate this previously unreported phenomenon as reactive glass-metal interaction (RGMI).

Post-ultrasonication surface analysis by FESEM and AFM reveals distinctive cavity formations in NAPS-X-Au glasses where GNIs were originally embedded, indicating strong mechanical anchoring of nanoislands to the substrate (Section S7; Figures 1A, S5 and S6). Significantly, NAPS-30-Au surfaces lack these characteristic depressions, suggesting substantially weaker interfacial bonding. Quantitative analysis demonstrates that cavity depth (Figure S6) systematically decreases with increasing $SiO_2$ concentration, revealing how compositional variations—particularly the $SiO_2:P_2O_5:Al_2O_3$ ratio—modulate the oxidative RGMI process and consequent nanoislands stability. This approach yields ultra-stable MEGNIs through a remarkably straightforward process that circumvents conventional requirements for chemical precursors, reducing agents, extreme temperatures, harsh reaction conditions, and extended synthesis protocols (*33–36*).

*Glass-Gold Interfacial Chemistry and Diffusion Profiles*

We employed X-ray photoelectron spectroscopy (XPS) depth profiling and angle-dependent XPS measurements (Schematics are shown in Figure 1C) to characterize chemical composition changes at glass surfaces following RGMI and to track GNI diffusion behavior. XPS depth profiles for NAPS-15-Au (Figure 1D) and NAPS-30-Au (Figures S7), representing integrated peak areas for each element (Figure S8) at varying depths, reveal gradual gold signal attenuation with depth. Notably, NAPS-30-Au exhibits a steeper gold intensity decline at shallower depths (Figure S7), indicating a more abrupt interfacial transition compared to NAPS-15-Au. To quantify GNI diffusion depth, we analyzed atomic fraction profiles for NAPS-15-Au (Figure 1E) and NAPS-30-Au (Figure S9), determining Au atomic fractions by normalizing depth-dependent intensities against standard Au foil. After calibrating sputtering yield using a $Si/SiO_2$ reference, we fitted the Au depth profiles using combined Fermi-Gauss functions:

$$I = I_0\{1 + \exp\left[\frac{d-d_{1/2}}{x}\right]\}^{-1} \quad (1)$$

$$I = \frac{1}{d_w\sqrt{2\pi}}\exp(-\frac{d^2}{2d_w^2}) \quad (2)$$

Where *I* represents quantified Au atomic fraction, $I_0$ is maximum signal intensity, *d* is sputter depth, *x* is Au intensity decrease rate, $d_{1/2}$ is sputter depth at half-maximum intensity, and $d_w$ represents interface width. Fitting results (Table S1) reveal broader interfaces and greater incorporation depths in NAPS-15-Au compared to NAPS-30-Au, reflecting compositional influences on thermal and physical properties.

Both samples display an initial gold intensity increase accompanied by sodium signal reduction—an effect more pronounced in NAPS-15-Au—highlighting $Na^+$ ion migration and its existence near or within GNIs. Comparative analysis of Au $4f_{7/2}$ spectra from these samples against standard Au foil (Figure S10) reveals subtle but significant spectral modifications, including peak broadening and binding energy shifts, potentially arising from Fermi-level position alterations (*37*). Measured full width at half maximum (FWHM) values (1.01 eV for NAPS-15-Au, 0.91 eV for NAPS-30-Au, versus 0.77 eV for standard Au) suggest new spectral components, indicating bidirectional migration where glass elements diffuse into GNIs while gold atoms simultaneously migrate into the glass matrix, transforming GNIs into MEGNIs.

Angle-dependent XPS measurements for NAPS-15-Au provide additional evidence of glass-GNI interactions. As shown in Figures 1F-1H, and S11 and Table S2, binding energy systematically shifts toward higher values with increasing analyzer angle—this effect is most pronounced for Au and P, less significant for Na, and entirely absent in standard Au foil. This phenomenon can be attributed to diminished core-hole relaxation effects as measurement depth increases, consistent with partial MEGNI embedding within the glass. The progressive decrease in Au and P signal intensities with increasing analyzer angle (Figure 1I) further supports element co-migration and depth-dependent compositional variation, underscoring the dynamic nature of RGMI and the critical role of glass-gold interfacial interactions in determining MEGNI structural and compositional evolution. These findings indicate MEGNIs possess a dual-region structure: (i) a core predominantly composed of Au atoms and (ii) a surface region enriched with both Au and glass-derived elements.

*Atomic-Scale Architecture of Multi-Element Gold Nano-Islands*

The nanostructure of MEGNIs extracted from NAPS-15-Au was characterized using high-resolution TEM (HRTEM). STEM-HAADF imaging (Figure 2A) reveals contrast variations across GNI surfaces, indicating compositional inhomogeneity resulting from glass-metal interactions. Bright-field HRTEM images (Figures 2B, 2C and S12) exhibit distinctive cascading features, confirming the presence of multiple interfaces within individual nanostructures. Quantitative analysis through inverse fast Fourier transformation (IFFT) (Figures 2C-2E, S13-S14) demonstrates significant lattice parameter variations from periphery to core. While standard Au(111) and Au(200) interplanar spacing (d-spacing) measure 0.235 nm and 0.20 nm respectively (*38*), the (111) lattice planes at GNI peripheries exhibit around 15% variations relative to core regions, indicating substantial stress/strain accommodation. This pronounced lattice dilation gradually relaxes toward nanoparticle centers, providing direct evidence of glass element incorporation at GNI boundaries. Based on d-spacing gradient analysis, we estimate glass element diffusion penetrates several nanometers into the GNIs (Figure 2E). Complementary elemental mapping in HAADF-STEM mode (Figure 1F, S15-S16) corroborates glass element migration into GNIs during thermal processing. Notably, this strain manifestation demonstrates crystallographic selectivity, with (111) planes exhibiting significant distortion while (200) planes remain relatively unperturbed throughout the nanostructure, suggesting preferential chemical interaction along specific crystallographic directions. Additional HRTEM analyses of GNIs extracted from other compositions appear in Figures S17 and S18.

To definitively determine whether GNIs exist as pure Au or as Au-rich solid solutions containing P, Na, and Al solutes, we acquired nanobeam diffraction (NBD) patterns along multiple zone axes (Figures 2G, 2H, S19). High-magnification HAADF images (Figures 2G(i), 2H(i), 2H(iii)) show GNIs oriented along [1-1-2] and [-111] zone axes, with corresponding NBD patterns (Figures 2G(ii), 2H(ii), 2H(iv)) obtained from designated "O" regions.

Quantitative histogram analysis (Figure S19) reveals 1-11 plane d-spacing of 0.218 nm—significantly contracted from the 0.235 nm spacing in pure Au. Similarly, 220 planes measured 0.131 nm versus 0.144 nm in pure Au. These systematic lattice contractions relative to pure Au 002 and 112 spacing provide compelling evidence for Au-rich solid solution formation, consistent with incorporation of smaller-radius P, Si and Al atoms within the Au lattice. Additional histogram analysis of patterns in Figures 2H(ii) and 2H(iv) confirms 02-2 plane spacing of 0.131 nm and 0.130 nm, further validating lattice contraction and solid solution formation. The remarkably consistent d-spacing measurements across different regions within individual particles (Figure 2H(i,iii)) demonstrate homogeneous solute distribution throughout the Au-rich phase.

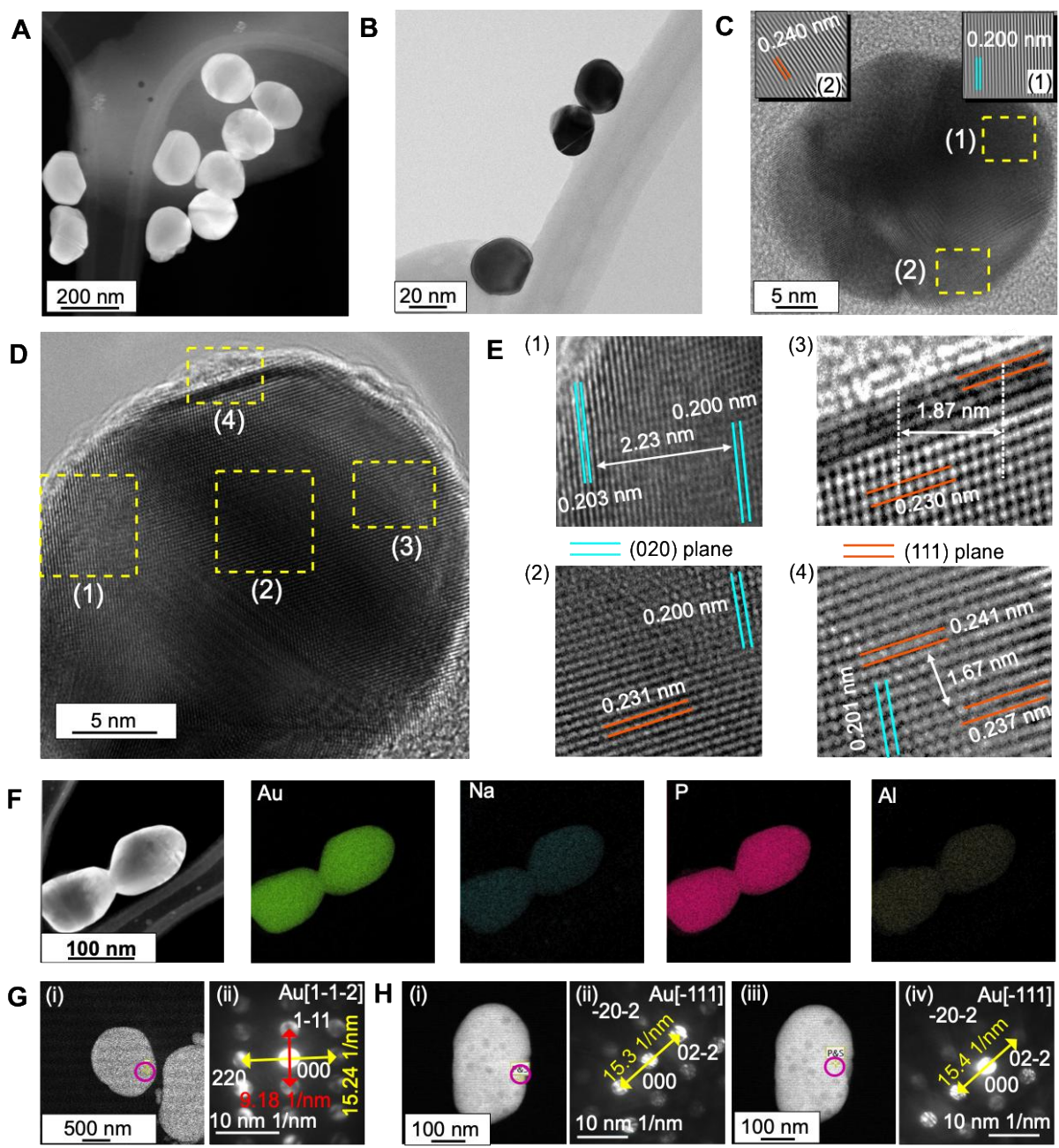

**Figure 2: High-resolution structural and elemental imaging of Au nano-islands.** *(A) Low resolution Dark field STEM-HAADF image. (B) Bright field TEM image. (C) and (D) High resolution TEM image for GNIs extracted from NAPS-15-Au. (E) Inverse fast Fourier*

*transformation (IFFT) images for the regions highlighted in Figure 2D. (F) Dark field HAADF-STEM and Au, Na, P and Al elemental mapping images for extracted MEGNI from NAPS-15-Au. (G) (i) The high magnification HAADF image of GNI. (ii) The nanobeam diffraction pattern (NBD) acquired along the Au[1-1-2] zone axis from the "O" mark region of Au nanoparticle shown in (i). (H) (i) and (iii) the high magnification HAADF image of GNI. (ii) and (iv) The nanobeam diffraction pattern (NBD) acquired along the Au[-111] zone axis from the "O" mark region of Au nanoparticle shown in Figure (i) and (iii).*

*Mechanistic Insights into Reactive Glass-Metal Interactions*

Despite prevailing assumptions of weak interactions between Au and oxide supports due to gold's low surface energy, oxygen-mediated SMSI (O-SMSI) have been documented between gold nanoparticles and phosphate-based materials, including hydroxyapatite ($Ca_{10}(PO_4)_6(OH)_2$) (*39*), $LaPO_4$ (*40*) and $AlPO_4$ (*41*). These observations suggest phosphate ($PO_4^{3-}$) anions critically facilitate metal nanoparticle encapsulation (*31, 42*). Our molecular dynamics (MD) simulations of NAPS-X glass structures confirm the presence of $PO_4^{3-}$ species predominantly connected to $Na^+$ and $Al^{3+}$ (*43*)—indicating potential $Na^+$–$PO_4^{3-}$ and $Al^{3+}$–$PO_4^{3-}$ mediated interactions with GNIs at elevated temperatures. The MD simulations (Figure 3A) reveal systematic structural evolution with increasing silica concentration: Si–O–Al and Si–O–Na linkages proliferate while P–O–Al connectivity diminishes. This compositional shift produces progressively more compact glass networks, evidenced by decreasing molar volume ($V_m$) (Figure 3B). Notably, $Na^+$ cations exhibit the highest diffusion coefficients among glass constituents (Figures 3C and S20), with mobility increasing exponentially with temperature (Figure 3C inset).

To quantify gold diffusion dynamics, we deposited 50 nm-thick Au films on multiple NAPS-15 substrates and performed thermal treatments at 300 °C, 375 °C, 450 °C, 500 °C, and 550 °C for 15 minutes in air. AFM and FESEM analysis (Figure 3D (bottom)) reveal GNI mean height (Figure 3E) initially increases from 300 °C to 450 °C, followed by pronounced reduction at higher temperatures. Similar patterns emerged across unique substrate (Figure 3D (top)) (See section S11). The sharp decrease in average GNI height near glass transition temperature ($T_g$) (Figure 3B and Figure S21A) correlates with the temperature-dependent $V_m$ expansion (Figure S22) and strong correlation among $V_m$, $T_g$ and thermal expansion coefficient ($\alpha$) (Figure 3B inset, Figure S21B and S21C), underscoring the critical role of $V_m$ and $T_g$ in governing gold diffusion behavior. However, since NAPS-30 exhibits a lower $T_g$ (443 °C) than NAPS-15 (472 °C) (Figure 3B), yet shows weaker GNI-glass interaction (Figure S5 and S6), $T_g$ alone cannot fully account for the observed behaviour. This suggests that the diffusion dynamics of GNIs are governed by a combined effect of $V_m$, $T_g$, intrinsic chemical composition of the glass, and, more importantly, the distribution of structural units.

Our thermal processing temperature (550 °C) substantially exceeds gold's Tammann temperature (~400 °C)—the threshold for significant atomic mobility. In glass networks, $Na^+$ migration proceeds via hopping mechanisms between adjacent cationic sites. XPS depth profiles for NAPS-15(30)-Au reveal $Na^+$ concentration gradients, suggesting outward migration. These liberated $Na^+$ ions, upon leaving their oxygen coordination shells, likely interact electrostatically with gold atoms to form Au–Na bonds (*44*). Concurrent $Na^+$-Au interdiffusion processes resembling $Ag^+$–$Na^+$ ion exchange mechanisms cannot be discounted (*45*). The resulting $Na^+$ depletion and Au substitution in the glass network facilitates reorientation and adsorption of phosphate, aluminate, and silicate anions onto GNI surfaces via oxygen-mediated bridges (*46*). Given $Au_2O_3$ decomposition occurs around 300 °C, oxygen evolution from near-surface regions is expected, consistent with XPS depth profiles showing characteristic S-shaped oxygen distribution curves with dual diffusion tails at low and high concentrations (Figure 1D). This pattern suggests GNIs actively participate in oxygen release at NAPS-15-Au and NAPS-30-Au surfaces, indicating catalytic activity. Surface depressions observed in all NAPS-X glasses except NAPS-30-Au (Figures S5 and S6) directly beneath nanoparticles (*15*) provide additional evidence for catalytic interaction between GNIs and glass substrates during thermal processing (*47*).

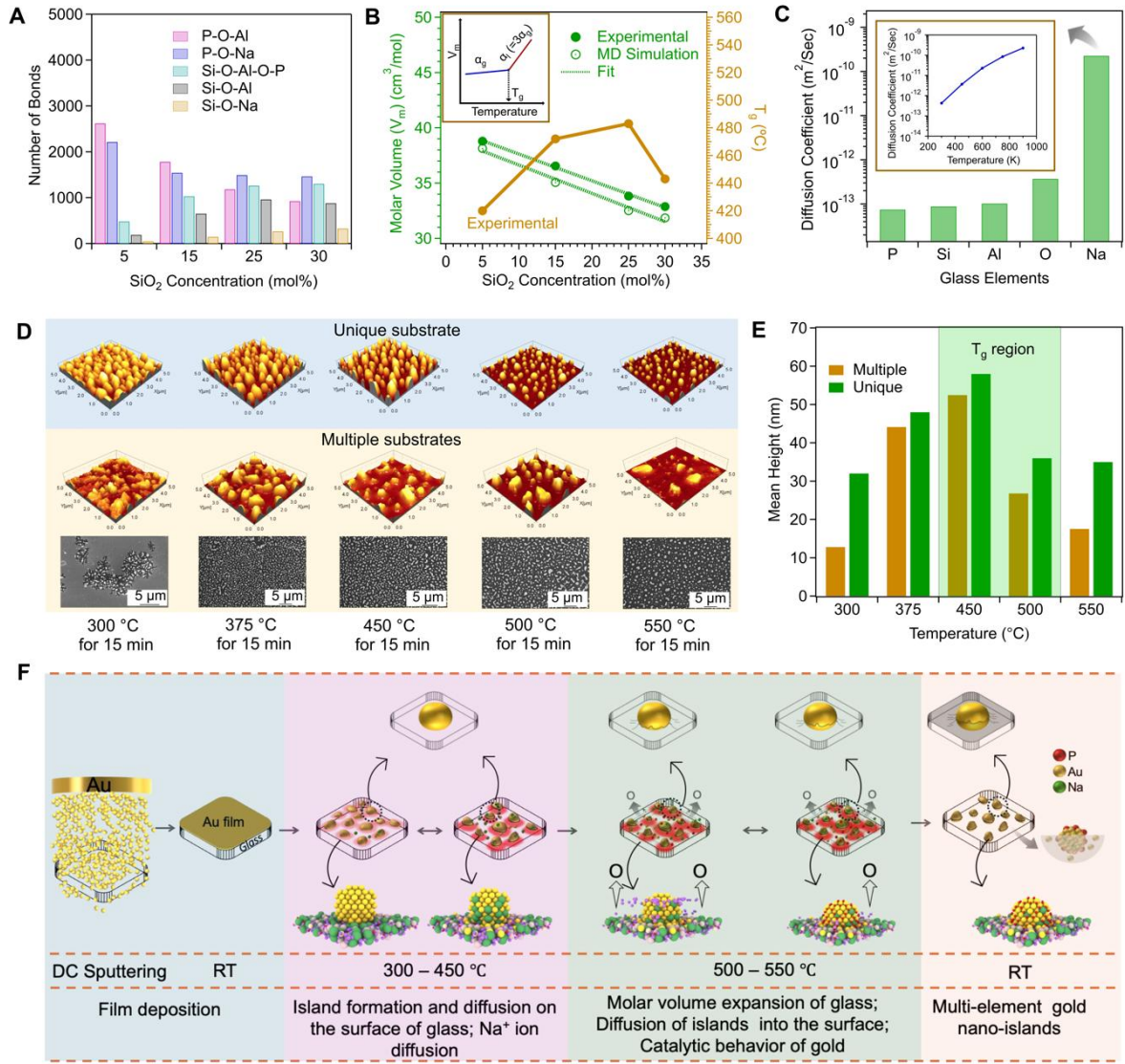

**Figure 3: Mechanism behind RGMI.** *(A) Variations in the number of P-O-Al, Si-O-Al, P-O-Al-O-Si, Si-O-Na, and P-O-Na bonds with incrasing the SiO₂ concentrations in NAPS-X glasses. Inset (i) shows the Si-O-Al and P-O-Al bond angle distribution for NAPS-15, measured using MD simulations. (B) Molar volume ($V_m$) and glass transition temperature ($T_g$) variation with increasing the SiO₂ concentration in NAPS-X glasses. Inset shows the schematic picture representing the changes in molar volume with increasing the temperature of glass. $\alpha_g$ and $\alpha_l$ represents the volume thermal expansion coefficient of glass and liquids. (C) Diffusion coefficient for elements exist in NAPS-15-Au glass at 900 K. Inset shows the variation in Diffusion coefficient with tempeature for Na. (D) (top) AFM morphology of sputtered gold on unique substrate followed by sequential heat treatment at different temperature. (bottom) AFM and FESEM morphology of sputtered gold on different substrates at different temperature. (E) Mean height variation of HGNIs with increasing the temperature. Mean heights were*

*calculated using AFM morphological images shown in Figure 3D. (F) Schematic picture representing the formation of MEGNIs using the NAPS glasses.*

Figure 3F schematically illustrates the RGMI mechanism. Unlike conventional RMSI processes in reducing environments (described by $Au + yMO_x + xyH_2 \rightarrow AuM_y + xyH_2O$, where $MO_x$ represents reducible oxides), our RMGI phenomenon occurs with non-reducible NAPS glasses under oxidizing conditions. $Na^+$ diffusion and Au-Na exchange facilitate interactons between gold and P–O–Al/P–O–Na moieties. The catalytic activity of GNIs promotes oxygen evolution and enhances $Na^+$, Al and P incorporation. The dominant reaction pathway follows $Au + yBO_x \rightarrow AuB_y + xyO_2\uparrow$, where B represents P, Si and Al (*48*). We conclude that multiple factors—deposition technique (Figure S23), $V_m$, $T_g$, $Na^+$ mobility, network connectivity among $PO_4$, $AlO_4$, $SiO_4$, and $NaO_5$ structural units, and gold's catalytic properties—collectively drive the RGMI process. The inherent challenges of in-situ characterization in oxidizing atmospheres (Figure S24) necessitate further investigations to definitively identify dominant diffusion mechanisms and most mobile species. (see the Section S10 and S11 for further explanation)

*Investigating Electronic Structure Modifications in MEGNIs*

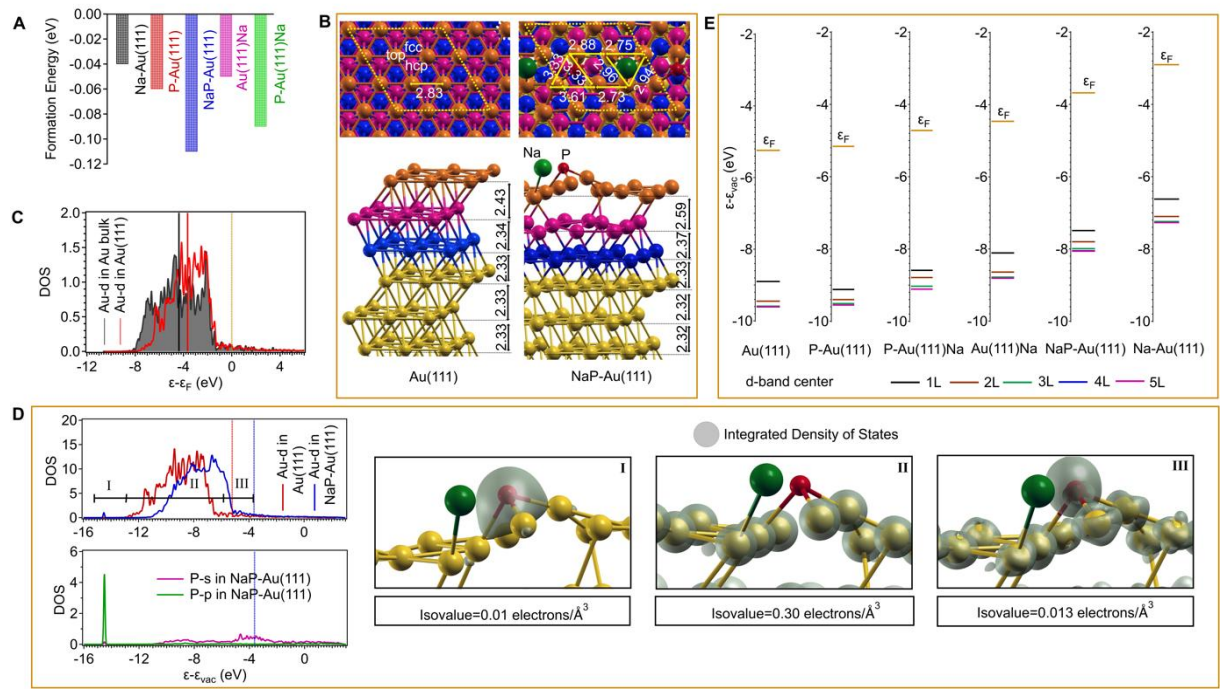

**Figure 4: Density functional theory.** (*A*) *Energy bar diagram comparing the formation energies of different Au(111) slab models. (B) Top and side view of clean Au(111) and NaP-Au(111). The figure marks the three adsorption sites (top, hcp, fcc). For better understanding of the sites we have colored Au in the top three layers in different color. In the remaining layers Au are denoted with a golden color. Na and P are denoted with green and red colors, respectively. (C) DOS of Au-d in bulk and top layer of Au(111). All energies are shifted with respect to Fermi energy (orange bold line). Black and red bold lines denote d-band centers of Au-d for bulk and Au(111), respectively. (D) DOS of Au-d and P-s,p in NaP-Au(111). DOS of Au-d in Au(111) is shown as a shaded region for comparison. Energies are shifted with respect to vacuum energy. The vertical dotted lines denote the Fermi energy for Au(111) (red) and NaP-Au(111) (blue). Integrated DOS are shown beside the figure for the three regions marked*

*in the plot. (E) d-band centers for different Au(111) slab models of Au in top five layers (see methods is SI).*

Based on compositional gradients observed via XPS (Figure 1) and structural perturbations identified through HRTEM (Figure 2), we performed plane-wave density functional theory (DFT) calculations to systematically investigate dopant effects on Au(111) surface properties. We modeled five distinct systems: Model 1 – Na adsorbed on Au(111) (Na-Au(111)); Model 2 – P adsorbed on Au(111) (P-Au(111)); Model 3 – Na substitution in Au(111) (Au(111)Na); Model 4 – P adsorbed on Na-Au(111) (NaP-Au(111)) and Model 5 – P adsorbed on Au(111)Na (P-Au(111)Na). Our energy analysis (see methods in SI) (Figure 4A) indicates P adsorption on Na-Au(111), i.e. Model 4, is thermodynamically favored. Nonetheless, the formation of multiple doped surface structures at elevated temperatures remains plausible. From our comprehensive analysis of five different Au(111) slab models with Na and P dopants (see Section S12), we focus here on NaP-Au(111) configuration that best elucidate Au-Na and Au-P interactions.

Figure 4B illustrates the characteristic A-B-C stacking sequence of pristine Au(111) and NaP-Au(111). Among the three high-symmetry adsorption sites on Au(111) (Figure 4B), Na and P are adsorbed at hcp (hollow site formed straight above the 'B' layer) and fcc (hollow site formed straight above the 'C' layer) sites, respectively. The top view structure notes the variation in bond length around the doping sites. Bond length calculations (Table S3) for NaP-Au(111) slab reveal the following observations: Na-P distances in one direction is around 4.28 Å while in another direction is around 7.25 Å; Na–Na and P–P distances are consistently ~8.50 Å in both directions. The Na–Au bond length is ~3.07 Å, slightly longer than in other doped Au(111) configurations, indicating a weaker interaction. In contrast, the P–Au bond length remains relatively unchanged at ~2.35 Å across all models. Structural analysis across the series of doped slabs reveals that the simultaneous presence of Na and P induces pronounced lattice distortions not only in the topmost atomic layer of Au(111) but also in the subsurface (second) layer. As a consequence we find noticeable increase in the interplanar distances in NaP-Au(111) which are also noted in Figure 4B and S25. This structural response aligns well with the interlayer expansion observed in the corresponding TEM analysis (Figure 2), providing further validation of the modelled configurations.

Density of states (DOS) analysis comparing bulk and surface Au *d*-orbitals (Figure 4C) reveals quantum confinement effects perpendicular to the surface, manifested as narrowed d-band width for surface atoms relative to bulk. Consequently, the surface Au d-band center shifts approximately 0.6 eV higher in energy compared to bulk Au. Figure 4D shows modification in the DOS for surface Au-*d* states as an effect of adsorption of Na and P on Au(111). The DOS profile for NaP-Au(111) exhibits significant upward shift of around 1.5 eV in Au d-states, accompanied by a new electronic state at approximately –14.5 eV arising from P-*s,p* and Au-*d* orbital hybridization (Figure 4D). The Na *s,p* states, which lie much lower in energy (outside the plotted range), suggest minimal orbital overlap with Au, consistent with the lack of electronic interaction observed in angle-resolved Na 1s XPS spectra. These results indicate that Au–Na interactions in NaP–Au(111) are primarily electrostatic in nature. However, P-*s* interacts with Au-*d*, specifically Au-$d_{x2-y2}$ and Au-$d_{xy}$, near the Fermi energy. Moreover, P-*p* hybridizes with $t_{2g}$ states ($d_{zx}$, $d_{zy}$ and $d_{xy}$) of Au-*d*. For better elucidation of these interactions, we split the energy range into three parts and determined integrated DOS, which are shown in Figure 4D.

Analysis of d-band center (see methods in SI) positions across the top five atomic layers (Figure 4E) reveals a constant shift in the d-band center for Na-Au(111) and Au(111)Na relative to clean Au(111)—consistent with the purely electrostatic Na-Au interaction producing a constant potential shift ($\Delta\varepsilon$). In contrast, other doped slabs show differential shifts, exhibiting 0.1-0.15 eV upward shifts relative to corresponding layers in pristine Au(111). Comparative analysis across all five model systems demonstrates that while Na adsorption produces primarily electrostatic effects, P adsorption fundamentally modifies electronic structure. Furthermore, the emergence of P *s*-states near the Fermi level ($\varepsilon_F$) (Figure 4D) potentially creates reactive sites that facilitate charge transfer to adsorbate molecules, enhancing the catalytic performance of MEGNIs (*49*).

*Ultrafast Carrier Dynamics in MEGNIs*

In plasmonic metal nanoparticles, photoexcitation or plasmon decay generates non-equilibrium "hot" carriers with electron temperatures ($T_e$) substantially exceeding lattice temperatures ($T_l$) (Figure 5A; Scheme S2). These energetic carriers typically dissipate energy through sequential electron-phonon (e-ph) coupling followed by phonon-phonon (ph-ph) interactions with the surrounding medium, occurring on femtosecond-to-picosecond timescales. While this energy dissipation pathway dominates in isolated systems, the technological significance of plasmonic nanostructures for photodetection and photocatalysis hinges on efficient hot carrier injection into adjacent semiconductors or adsorbed molecules. In our multi-element glass-gold systems, the complex interfaces with varying lattice spacings and modified electronic density of states near the Fermi level present unique opportunities to modulate hot carrier dynamics (*50–52*), with direct implications for catalytic and sensing performance (*53–56*).

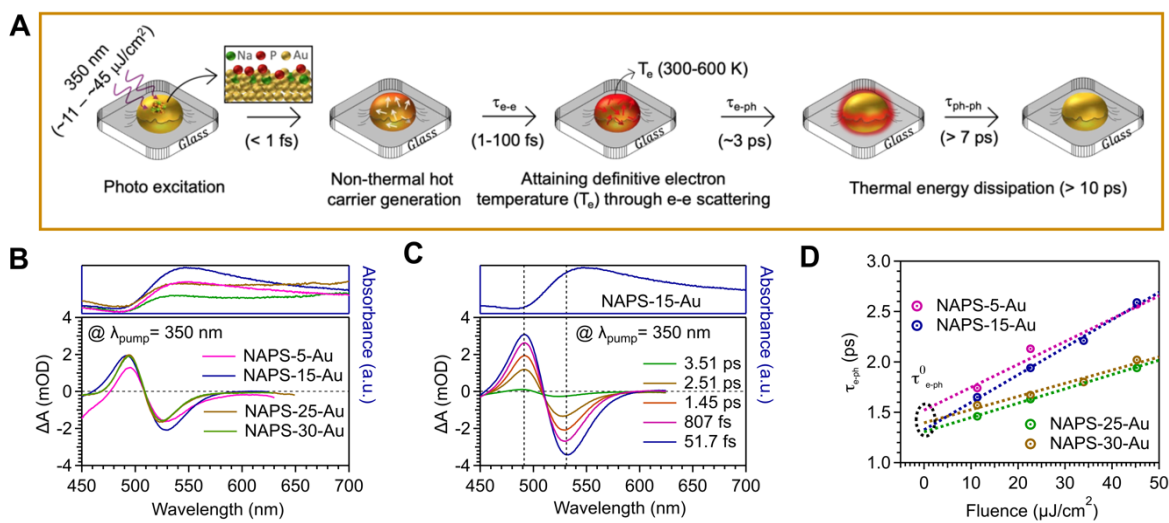

**Figure 5: Hot carrier dynamics.** *(A) Schematic of the hot electron formation and their relaxation in MEGNIs supported by the glass materials (see the Scheme S2 in supporting). (B) Comparison of transient absorption spectra for NAPS-X-Au glasses at 1.47 ps delay time. The steady state absorption spectra for all the NAPS-X-Au glasses is shown in the top panel of the figure. (C) Transient absorption spectra for NAPS-15-Au glass at different delay time. (D) Electron phonon coupling time for NAPS-X-Au glasses with increasing the pump fluence.*

Ultrafast transient absorption (TA) measurements across the 450-800 nm spectral range using a 350 nm pump at 1.45 ps delay (Figure 5B) reveal characteristic negative differential absorption features centered at ~530 nm for all NAPS-X-Au samples, corresponding to ground state bleaching (GSB) of the LSPR (*57*). The GSB spectral width and amplitude—proportional to nanoparticle dimensions at constant excitation power—systematically decrease with increasing silica content, indicating progressively larger GNIs with attenuated plasmonic response (Figure 5B, S26-S28) (*58*). Time-resolved TA spectra for NAPS-15-Au (Figure 5C) and corresponding bleach recovery kinetics (Figure S29) fitted with bi-exponential functions yield electron-phonon ($\tau_{e-ph}$) and phonon-phonon ($\tau_{ph-ph}$) coupling time constants. Notably, $\tau_{e-ph}$ values (Figure 5D) at equivalent pump powers are consistently higher for NAPS-05-Au and NAPS-15-Au compared to NAPS-25-Au and NAPS-30-Au, correlating with their enhanced mechanical embedding within the glass matrix. These observations confirm that hot electron dynamics in MEGNIs are governed by the surrounding dielectric environment, degree of embedment, and interfacial stability (*59*). The electron-phonon coupling constants (*g*) (see section S13), which remains temperature-independent throughout our experimental range (pump powers ≤ 800 μW, $T_e$ ≤ 600 K, Figure S29 and S30), increases systematically with silica concentration up to 25 mol% before decreasing for NAPS-30-Au. We attribute these variations to composition-dependent modifications in gold's electronic structure near the Fermi level, directly reflecting the extent of gold-glass chemical interactions (*60*). The combination of well-defined LSPR features, efficient hot carrier generation with elevated electron temperatures, extended relaxation lifetimes, and pump-wavelength-dependent optical responses (Figure S28 and S29) positions our ultra-stable MEGNIs—particularly those in NAPS-15-Au—as promising platforms for next-generation catalytic, photodetection, and sensing applications (*61–63*).

**Conclusions:**

Our work establishes RMGI as an effective strategy for engineering GNIs with enhanced stability and tailored surface chemistry. By manipulating sodium aluminophosphosilicate (NAPS) glass composition, we demonstrate that glass chemistry directly modulates interfacial phenomena, enabling gold nanostructure stability at moderate temperatures (550 °C) in ambient conditions. This approach bypasses requirements for chemical precursors, reducing agents, extreme temperatures, and extended reaction times, offering an environmentally sustainable fabrication pathway.

Density functional theory calculations illuminate the RGMI mechanism: Na adsorption at the outermost Au(111) layer through electrostatic interactions maintain gold's electronic structure, while phosphorus adsorbs at FCC hollow sites, inducing structural distortions and generating new electronic states through P-Au orbital hybridization. These modifications, particularly P p-states near the Fermi level, create catalytic sites that enhance charge transfer processes critical for surface reactivity.

MEGNI stability derives from complementary factors: network connectivity among phosphate, aluminate, silicate, and sodium oxide units; composition-dependent thermal behavior near the glass transition; and gold's catalytic activity facilitating oxygen evolution and interfacial rearrangements. Ultrafast spectroscopy confirms that modified surface states directly influence hot carrier dynamics, providing pathways to engineer energy transfer for enhanced photocatalytic performance.

The NAPS glass platform enables controlled chemical interactions with gold metal, establishing a versatile approach for thermally robust, multi-element nanostructures with tunable surface properties. These engineered materials advance applications in heterogeneous catalysis, optoelectronic devices, chemical sensing, and energy conversion.

**Acknowledgement:** This work was developed under the frame of the project funded by Council of Scientific and Industrial Research (CSIR), India through Fundamental and Innovative Research in Science of Tomorrow (FIRST) (Ref: FIR060301). Part of the work was performed under the frame of project support by the Science and Engineering Research Board (SERB)/Anusandhan National Research Foundation (ANRF), Government of India, India (CRG/2022/ 000930). SH and ARA acknowledges the CSIR, India and ANRF, India for financial support. S.M. thanks MHRD (Govt. of India) for the Prime Minister's Research Fellowship (PMRF ID: 1401240). P.K.C. thanks the Department of Science and Technology, DST-FIST (SR/FST/CS-II-027/2014), New Delhi, India, for providing funds for the ultrafast TA facility. S.S. express their gratitude to the Gdańsk University of Technology for their support under the "Excellence Initiative Research University (IUDB)" program. The computations were carried out using High Performance computing facilities of the Center of Informatics at the Tricity Academic Supercomputer & Network (Poland) and Poland's High-Performance Infrastructure PLGrid (ACK Cyfronet Ares). N. J would like to acknowledge CDAC Pune for providing computational facility. N. J would also like to extend sincere gratitude to Dr. T. Raja, CSIR National Chemical Laboratory, Pune for fellowship and providing computational facility. SS and NJ would like to thank Dr. Prasenjit Ghosh from IISER Pune for his valuable insights. EV acknowledge the Department of Science and Technology (DST), New Delhi Government of India, sanction order INT/RUS/RFBR/386 for partial financial support and SRM IST, KTR campus for the research infrastructure. PhD Scholar Mr Abimannan S is acknowledged for his help in handling samples. EV and ARA acknowledge Dr. Ambuj Mishra and Dr. Debdulal Kabiraj, Inter-University Accelerator Centre (IUAC) for performing the partial TEM measurements.

# Supplementary Information

# Reactive Glass–Metal Interaction under Ambient Conditions Enables Surface Modification of Gold Nano-islands

Sinorul Haque,[a,b#] Shweta R. Keshri,[c,#] G. Ganesh,[d] Kaustuv Chatterjee,[a,b] Shubhangi Majumdar,[e] Sudheer Ganisetti,[f] Indrajeet Mandal,[g] Dudekula Althaf Basha,[h] Prabir Pal,[a,b] Pramit K Chowdhury,[e] Niharika Joshi,[i] Subrahmanyam Sappati,[j] Nitya Nand Gosvami,[c] Eswaraiah Varrla,[d*] N. M. Anoop Krishnan,[f*] Amarnath R. Allu,[a,b*]

[a] CSIR-Central Glass and Ceramic Research Institute, Kolkata, 700032, India
[b] Academy of Scientific and Innovative Research (AcSIR), Ghaziabad 201002, India
[c] Department of Materials Science and Engineering, Indian Institute of Technology Delhi, Hauz Khas, New Delhi 110016, India
[d] Sustainable Nanomaterials and Technologies Lab, Department of Physics and Nanotechnology, SRM Institute of Science and Technology, Kattankulathur, Chengalpattu, Tamil Nadu 603203, India
[e] Department of Chemistry, Indian Institute of Technology Delhi, Hauz Khas, New Delhi 110016, India
[f] Department of Civil Engineering, Indian Institute of Technology Delhi, New Delhi 110016, India
[g] School of Interdisciplinary Research, Indian Institute of Technology Delhi, Hauz Khas, New Delhi 110016, India
[h] Department of Metallurgy Engineering and Materials Science, Indian Institute of Technology Indore, Simrol, Indore, 453552, India
[i] Catalysis and Inorganic Chemistry Division, CSIR National Chemical Laboratory, Pune-411008, India
[j] BioTechMed Center, and Department of Pharmaceutical Technologyand Biochemistry, ul. Narutowicza 11/12 Gdańsk University of Technology, 80-233 Gdańsk, Poland

# These authors contributed equally to this work
**Corresponding author:**
E-mail:
eswarail@srmist.edu.in (**Eswaraiah Varrla**)
krishnan@iitd.ac.in (**N.M.A. Krishnan**)
alluamarnathreddy@gmail.com; aareddy@cgcri.res.in (**A. R. Allu**)
Fax: +91-33-24730957; Tel: +91-33-23223370

## S1: Materials and Methods
*Synthesis of glasses*

The glasses used in the present work having the chemical composition $37.5Na_2O–25Al_2O_3–(37.5-x)P_2O_5–xSiO_2$ where x= 5, and 15 (NAPS-5 and NAPS-15 respectively) and $37.5Na_2O–(25-x)Al_2O_3–17.5P_2O_5–(20+x)SiO_2$ where x= 5, and 10 (NAPS-25 and NAPS-30 respectively) were synthesized using the standard melt quenching technique as proposed elsewhere (*1*). After the batch calculation, appropriate amounts of high-quality raw materials; sodium phosphate ($Na_3PO_4$, 96%, Sigma-Aldrich), ammonium phosphate monobasic ($NH_6PO_4$, ACS reagent, ≥98%, Sigma-Aldrich), aluminium oxide ($Al_2O_3$, powder, primarily α phase, ≤ 10 μm avg. part. Size, 99.5% trace metal basis, Sigma-Aldrich) and silicon dioxide ($SiO_2$, ~325 mesh, 99.5% trace metal basis, Sigma-Aldrich) were ground in a mortar-pestle to attain uniformity and homogeneity, followed by melting at 1250-1550 °C for 1 hour. The melt-quenched samples were annealed at $0.9T_g$ to remove the residual stresses generated upon quenching. At last, the samples were cut and polished with various grits (320-4000P) of SiC paper to achieve an optical grade finish and transparency.

### *Deposition of gold film through DC sputtering*

NAPS-X glass slides with dimensions of 10 mm × 10 mm × 1.5 mm were used for depositing a 50 nm thick gold (Au) film. After cleaning, the slides were loaded into an Agar Auto Sputter Coater (Model: 108A, UK). The thickness of the deposited film was monitored using an Agar Thickness Monitor (Model: MTM–10).

### *XRD*

The crystalline nature of the Au-coated (NAPS-X-Au) samples were analysed using a Rigaku Ultima 4 X-Ray Diffractometer using Cu $K_α$ anode with a scan rate of 2°/min. Furthermore, the NAPS-15-Au samples were heat-treated at various temperatures (300, 375, 450, 500 and 550 °C for 15 minutes) and their XRD measurements were taken.

### *UV-VIS-Absorption*

Optical absorption spectra for NAPS-X-Au glasses were recorded in the wavelength range of 300−800 nm using a Thermo Scientific UV–vis absorption spectrometer (Model: Evolution 200, US).

### *Ultrasonication*

Ultrasonication using a bath sonicator (Barson 1800, frequency of 40 kHz, Power 90 W) was performed to extract the embedded gold nano-islands (GNIs) from the NAPS-X-Au glasses. Initially, the substrate was kept at the bottom of a 10 mL glass vial, and then 3 mL of acetone was added. The NAPS-X-Au glass with vial was bath sonicated for 3 h 30 min with different time intervals with the help of chilled water circulation. After this process, the substrate was removed from acetone. High-speed centrifugation was performed at 8000 rpm for 30 min to sediment the separated GNIs from the glass matrix in the solvent. After centrifugation, the top 99% of the acetone was removed entirely. It was observed that a light-red-colored sample dispersed in a slightly concentrated form, ≈100 μL of GNIs in acetone was preserved and used for further imaging analyses.

### *FESEM and EDS*

Thermo Fisher Scientific Apreo S FESEM instrument with an accelerating voltage of 10–30 kV in the high vacuum mode was used for capturing the SEM images. The elemental mapping, and line scans profiles for (Au, P, Na, O, Al, and Si were acquired using an inbuilt FESEM energy-dispersive x-ray (EDAX) detector (Ultra Dry Silicon Drift Detector). EDAX detectors with a specified resolution was 129 eV. High-angle annular dark-field, dark-field, and bright-field scanning transmission electron microscopy were acquired with ADF, STEM 3+ detectors. Secondary electron images were captured with an ETD detector. For STEM imaging, detached MEGNIs containing acetone solvent of≈15 μL drop-costed on the Lacey carbon-coated copper grid with a mesh size of 400. The grid was kept in a vacuum oven at 80 °C for 12 h before the microscopy experiment.

### *Atomic Force Microscopy (AFM)*

The nanoscale topography of the NAPS-X-Au samples were observed using a commercial Drive AFM (Nanosurf, Switzerland) in non-contact (tapping) mode. The topographical images were further analyzed using Gwyddion software to qualitatively estimate the Au nanoislands height distribution in each NAPS-X-Au sample. 3-4 topographical images were taken at different locations in the samples and the nanoislands were masked in each image.

These masked regions were then analyzed using the 'Measure individual grains' tool from the Gwyddion software to find the height of the nanoislands. The average height and standard deviation of the nanoislands were calculated using each of the data points. The variation in the height distribution of the Au nanoislands can be attributed to the compositional variation of the substrate glasses. Additionally, The NAPS-X-Au samples were subjected to ultrasonication leading to the evacuation of some of the Au nanoislands, thereby forming pits on the surface of the samples. These pits were further observed using AFM topographical image and were analyzed to find their depth distribution. The depth distribution analysis of each sample was performed to evaluate the depth until the Au nanoislands submerged on the different glasses upon heat treatment. Multiple line profiles were taken on each pits and the depth of the pits were evaluated. The average depths of the Au nanoislands on each of the glasses were calculated by averaging the data from all the nanoisland pits. Moreover, NAPS-15-Au samples were heat treated individually at 300, 375, 450, 500 and 550 °C for 15 min and their topographical images were taken using AFM. The evolution of GNIs were intelligibly observed with increasing temperatures and their height distribution was calculated using Gwyddion software with the protocol mentioned before.

*XPS-depth profiling*

XPS measurements were performed in a commercial photoelectron spectrometer PHI 5000 Versa Probe-II under a base pressure of 3.7 x $10^{-10}$ mbar equipped with a monochromated Al K Alpha (1486.7 eV) focused X-ray source paired with a low-energy electron flood gun. The energy resolution for the core levels measurements was 400 meV taken at a pass energy 11.75 eV with a step size of 0.1 eV. Depth profile data was generated using a pulsed argon-ion sputter gun set at 1 kV accelerating voltage with a sputter area of 1mm x 1mm in conjunction with the X-ray source of 100 μm beam size keeping the analyzer pass energy set to 23.57 eV at a step size of 0.2 eV. The sputter yield for the sample was calculated to be around ~ 2 $A^0$/s by modifying the yield value obtained from standard $Si/SiO_2$. Charge correction for the insulating samples was done by calibrating the C 1s peak of the sample at 284.8 eV and the Au $4f_{7/2}$ peak of sputter-cleaned Au foil at 84.0 eV.

*Computational details for Density Functional Theory:*

We have used plane-wave based Quantum ESPRESSO software package for studying the clean and modified Au slab in order to understand the interesting properties observed at the GNI/glass interface (*2*). We have used ultrasoft pseudopotentials to describe the electron-ion interaction (*3*). The energy cutoffs for wavefunction and charge density are set as 60 and 360 Ry, respectively, such that the accuracy in energy is of the order of 1 meV. The electron-electron exchange-correlation potential is included using the parametrization developed by Perdew, Burke, and Ernzerhof (PBE) which uses generalized gradiant approximation (GGA) (*4*). An automatic Monkhorst Pack k-point grid of size $12 \times 12 \times 12$ per ($1 \times 1$) conventional unit cell of bulk Au is used for the integration of the Brillouin zone (*5*). Further, for speeding the calculations we have used Marzari-Vaderbilt smearing of width 0.005 Ry (*6*).

Further, we have modeled Au(111) slab for understanding the properties at the interface of GNIs and the glass support. The slab is 8 layers thick. The bottom 3 layers of the slab are fixed at the bulk parameters while the positions of atoms in the remaining 5 layers are optimized. In order to decrease the spurious interaction of the Au surface atoms with those at the periodic images of the slab, we have used a vacuum separation of thickness around 12 Å. Additionally, to eliminate the spurious long-range dipole interaction between the periodic images of the slab which is along the perpendicular direction to the surface we have included dipole correction. For predicting the interactions between the surface Au atoms and the dopants we have used an empirical van der Waal correction so as to include the dispersion interaction which are otherwise not taken into account by PBE-GGA functional (*7*).

For testing our pseudopotentials, we have performed calculations on Au bulk, Na bulk and black phosphorus. From our calculation we have determined the lattice parameters of bulk Au (FCC) to be around 4.17 Å with PBE-GGA functional and 4.01 Å using PBE-GGA functional corrected with the empirical van der Waals correction. The lattice parameter of bulk Na (BCC) was determined to be around 4.20 Å. For black phosphorus (orthorhombic) the optimized lattice parameters obtained with (without) van der Waals correction are: a=3.33(3.32) Å, b = 4.42(4.55) Å and c = 10.47(11.28) Å. Additionally, the buckling within a layer in black phosphorus is around 2.15 Å and P-P bondlength is around 2.23 Å. All these optimised structural properties for the three bulk systems determined from our calculations agree very well with previous reports (*8*). [*Mater. Res. Express,* **3,** 046501(2016)]

Determination of Formation energy:

$$E_{formation} = E_D - E_{Au} - (E_{Au(111)} + E_{Na+} + E_P) \quad (S1)$$

where, $E_D$ and $E_{Au(111)}$ are the total energies of doped and undoped Au(111) slab. $E_{Au}$ is the total energy of Au bulk per atom. $E_{Na}$ and $E_P$ are total energies of isolated Na and P.

Determination of *d*-band center:

$$d\text{-band center} = \left(\int_0^{\varepsilon_F} \varepsilon \cdot D(\varepsilon) \, d\varepsilon\right) / \left(\int_0^{\varepsilon_F} D(\varepsilon) \, d\varepsilon\right) \quad (S2)$$

where, $D(\varepsilon)$ is density of states at energy $\varepsilon$. $\varepsilon_F$ is Fermi energy.

Determining integrated density of states:

$$IDOS = \int_{\varepsilon_{min}}^{\varepsilon_{max}} D(\varepsilon) \, d\varepsilon \quad (S3)$$

where $D(\varepsilon)$ is density of states which is integrated between [$\varepsilon_{min}, \varepsilon_{max}$].

*TEM Sample preparation*

The nanoparticles are extracted from the NAPS-15 and NAPS-30 substrate using ultrasonication with a Branson 1800 bath sonicator at a frequency of 40 kHz and power of 90 watts. The substrate was placed in a 10 ml glass beaker containing 3.5 ml of acetone and subjected to 3 hours of bath sonication in an ice water bath, with 30-minute intervals. After sonication, the substrate was removed, and the acetone containing the nanoparticles was collected for further analysis. For TEM sample preparation, the recovered nanoparticles were diluted, and ~10 μl of the semi-transparent dispersion was drop-cast onto a lacey carbon copper grid (400 mesh). The grid was then dried in a vacuum oven at 80 °C for 12 hours.

*TEM Instrumentation*

JEOL, JEM 2100 Plus with an accelerated Voltage 200 kV in the high-vacuum mode, fitted with a single crystal of lanthanum hexa-boride (LaB$_6$) thermionic emission gun. Oxford X-Max EDAX detector was used for elemental analysis. Aperture size 50 μm (spot size 2 μm), objective lens aperture 60 μm and condenser lens aperture 150 μm. JEOL, JEM F200 with an accelerated Voltage 200 KV in high vacuum mode fitted with a with a cold field emission gun (CFEG). The Gatan in, Continuum NR Camera 1800GIFNR.STD used for Electron Energy Loss spectrum (EELS).

*Molecular dynamics simulations*

The simulated NAPS-X glass samples were prepared using molecular dynamics simulations via melt-quench approach using LAMMPS software package (*9*). The interaction between the atoms is modelled using Pedone potential (*10*). Periodic boundary conditions are imposed along all the three axes. The Newton's equation of motion is integrated using the Velocity Verlet algorithm with a timestep of 1 fs. The procedure followed to prepare the glass samples is similar to that in Ref. (*11*). Around 10000 atoms were initially placed randomly in a cubic box of length calculated by mimicking the experimental density. After relaxing the initial positions using molecular statics, the sample is thermalized for 50 ps at 300 K using micro-canonical ensemble (NVE). It was then heated from 300 K to 5000 K linearly at a rate of around 100 K/ps using canonical ensemble (NVT). In order to get a proper melt, the sample is held at 5000 K for 100 ps under NVT ensemble. The well equilibrated melt is then quenched at a rate of 1 K/ps until reaching 300 K under NVT ensemble. The stress accumulated during the quenching procedure because of constant volume is removed by allowing the atoms to relax at 300 K for 100 ps under NPT ensemble. A production run of 50 ps under NVE ensemble was performed for taking the statistical average. Bonding statistics, pair distributions, and angle distributions were computed every picosecond during the production run and the average is presented whenever necessary.

*DSC and $C_p$ measurements*

The DSC and $C_p$ measurements of the as-synthesized NAPS-X glasses were carried out using the Netzsch DSC 404 F3 Pegasus® instrument at 10Kpmin heating rate in the DSC Pt pans covered with lids with nitrogen gas flow of 60 and 40ml/min. All the measurements were recorded at the 2nd heating cycles, after the 1st heating and cooling cycle @10Kpmin, to ensure all samples possessing the same thermal history. Specifically, the sapphire standard specimen with a mass of ~57 mg was used for $C_p$ measurements, and the specific heat capacity of the comparable mass samples was evaluated using the $C_p$ ratio method between the standard specimen and the as-synthesized samples. The DSC and Cp measurements were further analyzed to define the glass-transition temperature ($T_g$) of the glasses.

*Transient Absorption Spectroscopy*

**Experimental Setup**: Transient absorption studies were performed at the Department of Chemistry, IIT Delhi. Ultrafast transient absorption spectral measurements were conducted using a commercial one-box ultrafast Ti: Sapphire amplifier (Astrella 1K-F, Coherent Inc.) pulses of 100 fs time duration, amplified pulse at 1 kHz repetition rate were obtained. The amplified pulses with 5 mJ/pulse energy were obtained after seeding with 20 femtosecond 70 nm bandwidth laser pulses obtained from an integrated oscillator (Vitara-S, 400 mW at 800 nm) pumped by Verdi-G, at 80 MHz repetition rate coupled to a femtosecond transient absorption spectrometer (Helios Fire UV-VIS, Ultrafast Systems). The amplified output of central wavelength 800 nm (5mJ/pulse, 1 kHz) was divided into two parts. One part (~200 mW) was used to produce the femtosecond probe pulse by focusing it either on a $CaF_2$ (for UV probe: 310 – 600 nm) and 2-mm thick sapphire crystal for generating white light continuum (400-800 nm) while the other part (3.25 mW) was used to generate a tunable femtosecond pump pulse using an optical parametric amplifier (Coherent OPeraASolo, 290-2600 nm). After the sample, the probe beam (1 kHz) was collimated and then focused into a fiber-optics coupled multichannel spectrometer equipped with CMOS sensor (1024 pixels) The pump power of the pump pulse used in the experiment was controlled by a variable neutral-density filter wheel and was kept in the range of 200-800 µW. Both the pump and probe beams were overlapped at the sample and the delay between the pump and probe pulses was controlled by a motorized delay stage. The pump pulses were chopped by a synchronized chopper at 500 Hz and the absorbance change was calculated with two time-adjacent probe pulses (pump-blocked and pump-unblocked).

**Experiments:** All the samples were held on the sample holder on constant translational motion at a speed of 3 mm/sec with a horizontal shift of 1.7 mm and vertical shift of 1 mm, within the area of (2 mm × 2 mm) during the measurements to avoid charring or damage of the samples. For transient measurements, 350 nm, 400 nm, and 450 nm were used as the pump wavelength, and sapphire crystal for visible range white light as the probe. Surface Xplorer version 4.5 was used to analyze the transient absorption spectra.

**Calculations of Decay Time:** The transient decays were fitted by using equation (S1), using a deconvoluting IRF of ~150 fs with the help of Surface Xplorer software and the averaged lifetimes were calculated according to equation (S2), where $\Delta A (\lambda, t)$ is the observed change in absorbance at time $t$ and wavelength $\lambda$, A is amplitude and $\tau_i$ is the time constant of $i^{th}$ components.

$$\Delta A (\lambda, t) = A_0 + \sum_i A_i\, e^{-\frac{t}{\tau_i}} \qquad (S1)$$

$$<\tau> = \frac{\sum_i A_i \tau_i}{\sum_i A_i} \qquad (S2)$$

## S2: Selection of glass compositions

Sodium aluminium phosphate (mol%) 37.5$Na_2O$-25.0 $Al_2O_3$-37.5$P_2O_5$ (NAPS-0) glass has been noted for its high ionic conductivity. To enhance the ionic conductivity of NAPS-0 glass to a range of $10^{-3}$ to $10^{-4}$ S/cm at room temperature, we added $SiO_2$ to the NAPS-0 glass. The impedance measurements reveal the high ionic conductivity of ~$10^{-4}$ S/cm at 250 °C for NAPS-0 glass containing $SiO_2$. Driven by curiosity, we intended to measure the ionic conductivity of later glass up to their glass transition range. We coated a 25 nm thick gold film as an electrode on the surfaces of glass for impedance measurements and cured the samples at around 550 °C for 1 hour. Interestingly, we have found that the coated gold film converted into nano-islands due to thermal dewetting phenomenon. In addition to that we have uncovered that glass substrate interact chemically with nano-islands and transformed in to firmly embedded hybrid gold nano-islands (HGNIs), which consist of multiple elements (Au, P, Al, Si, and Na), multiple oxidation states ($Au^+$, $Au^-$ and $Au^0$) and multiple interfaces. The formed HGNIs demonstrated superior performance in gas sensing and solar-driven water splitting. To the best of our knowledge, this is the first observation confirming chemical interaction between chemically inert glass and gold nanoparticles. However, the reasons behind this interaction remain unclear. Considering the wide range applications of multi element metal nano-particles it is therefore essential to find the reason behind the chemical interaction. To explore this further, we systematically varied the glass compositions by substituting $SiO_2$ for $P_2O_5$ and $SiO_2$ for $Al_2O_3$. This led to the design of five different glasses: NAPS-5 and NAPS-15 and NAPS-20, obtained by substituting 5 mol%, 15 mol%, and 20 mol% of $SiO_2$ for $P_2O_5$ in NAPS-0 glass, respectively; and NAPS-25 and NAPS-30, obtained by substituting 5 mol% and 10 mol% of $SiO_2$ for $Al_2O_3$ in (mol%) 37.5$Na_2O$-25.0$Al_2O_3$-17.5$P_2O_5$-20.0$SiO_2$ (NAPS-20). For the current study, we considered NAPS-5, NAPS-15, NAPS-25 and NAPS-30 glasses. Herein the glasses are named as NAPS-X, where X represents the total molar concentration of $SiO_2$. These glasses were able to synthesize using melt-quenching technique, a commercially viable method for synthesizing glasses in large scale.

## S3: Synthesis of firmly embedded gold Nano-islands on glasses

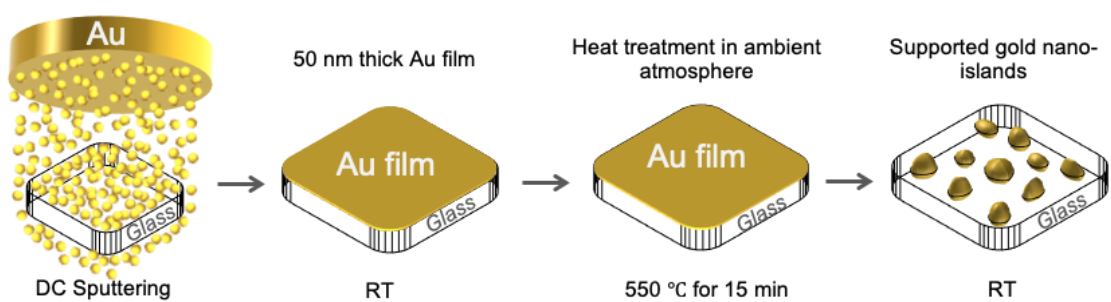

**Schematic S1**: Schematic picture demonstrating the synthesis of HGNIs on the surfaces of glasses.

Schematic S1 provides a detailed synthesis method of a commercially viable technique for preparing physically ultrastable GNIs on glass substrates. The specially designed NAPS-X glass enhances a chemical reaction between the deposited gold film and the glass during thermal treatment, which involves heating from room temperature (RT) to 550 °C at a heating rate of 150 °C/h, holding for 15 minutes at 550°C, and then allowing the system to cool naturally to RT. The entire synthesis process is conducted in an ambient atmosphere. Hereafter, NAPS-X glass with embedded GNIs are referred as NAPS-X-Au. For comparison purpose, 50 nm thick Au film was deposited on $SiO_2$ substrate and heat treated at 550 °C for 15 min under ambient atmosphere conditions. The substrate is named as $SiO_2$-Au.

### S4: Formation and morphology of gold nano-islands

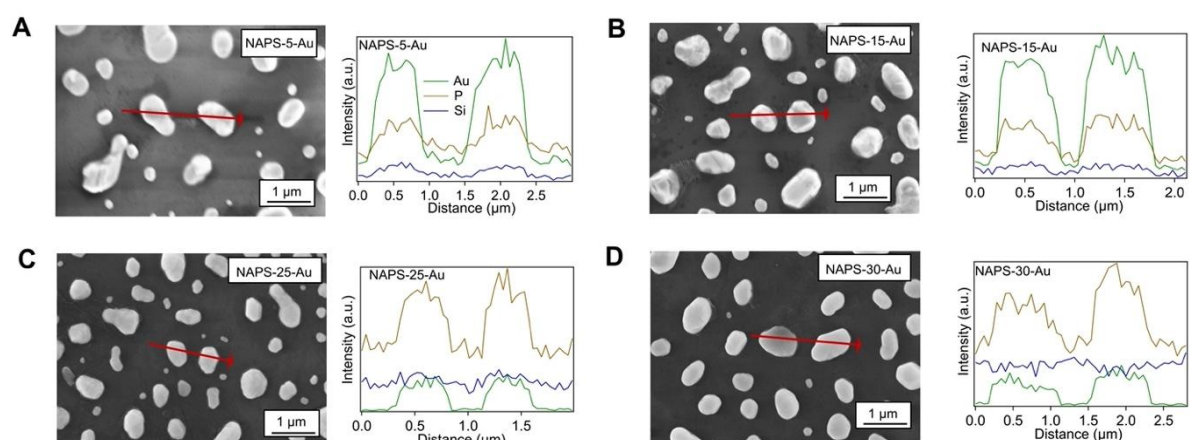

**Figure S1:** FESEM and EDS elemental line profile acros islands of HGNIs embedded in NAPS-X glasses.

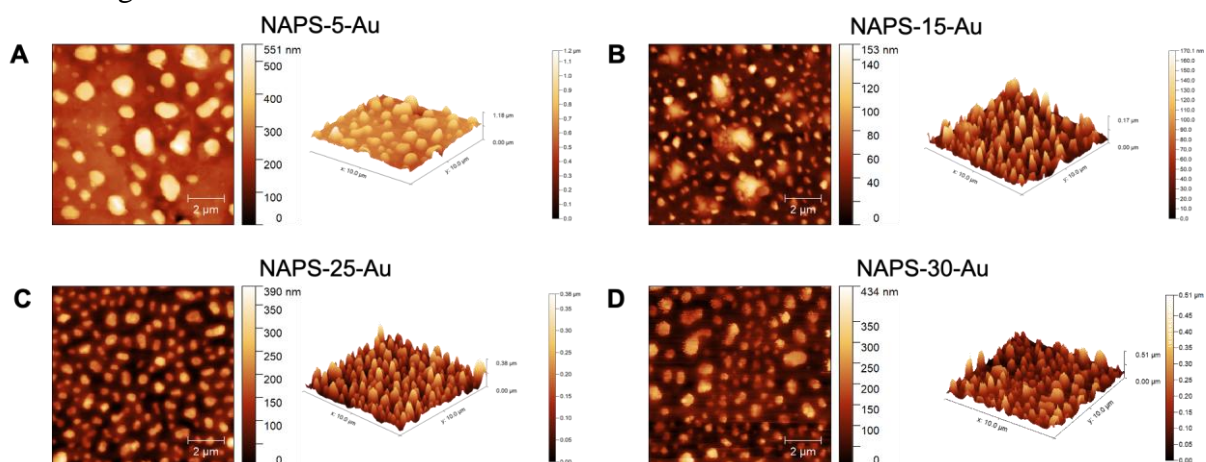

**Figure S2:** 2D and 3D AFM images of NAPS-X-Au glasses.

FESEM images with elemental (Si, P and Au) line profiles across the islands, and AFM images for NAPS-X-Au glasses, are shown in Figures S1 and S2, respectively. The 50 nm thick coated gold film has transformed into well-separated nano-islands. The GNIs on NAPS-5-Au, NAPS-15-Au, and NAPS-25-Au samples show significant coarsening, while those on NAPS-30-Au are much shinier and have smoother surfaces, indicating substantial alterations in the thermal dewetting process. Elemental line profiles confirm that the GNIs are enriched with phosphorus (P) and gold (Au). Importantly; the elemental line profile for silicon (Si) suggests a decrease in Si concentration on the surface of each island location with an increase in the SiO$_2$ molar concentration in NAPS-X glasses.

**S5: Structural and optical properties of embedded gold Nano-islands on NAPS-X glasses**

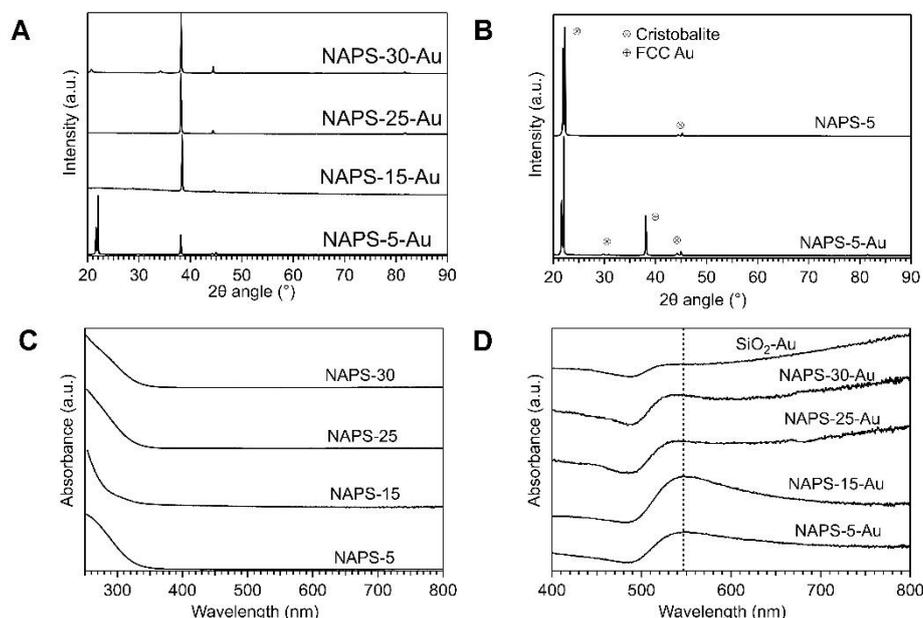

**Figure S3:** (A) XRD for NAPS-X-Au glasses and (B) comparision of NAPS-5 and NAPS-5-Au XRD spectra. Absorption spectra of (C) NAPS-X and (D) NAPS-X-Au glasses.

Figure S3A displays the X-ray diffraction (XRD) patterns for NAPS-X-Au glasses. The XRD patterns for NAPS-15-Au, NAPS-25-Au, and NAPS-30-Au show intense sharp peaks at a 2θ angle of 38.1°, with additional lower-intensity peaks at 44.3°, 64.6°, and 77.4°. These peaks correspond to the (111), (200), (220), and (311) planes of the face-centered cubic (FCC) lattice of gold (Au). For NAPS-5-Au glass, the XRD peak at a 2θ of 22° dominates over the peaks associated with FCC gold. Further, to confirm the influence of GNIs on the formation of crystalline phases, we heat treated NAPS-5 glass at 550 °C for 15 min. NAPS-5 exhibit XRD peaks (Figure S3B) similar to NAPS-5-Au except FCC Au peaks. The peak at 22° is attributed to the cristobalite crystalline phase, a polymorph of SiO$_2$ that often forms in silicate glasses with a high SiO$_2$ content at elevated temperatures. The cristobalite phase is purely composed of Si-O-Si linkages. Molecular dynamics simulations for NAPS-5 glass confirms that Si-O-Al and Si-O-Al-O-P bonds dominate over Si-O-Si bonds. The formation of cristobalite at relatively low temperatures and within the short heat treatment time of 15 min at 550°C confirms the network structural rearrangement in NAPS-5 glass with increasing heat treatment (*12*).

Figure S3C and S3D presents the absorption spectra for NAPS-X-Au glasses. All NAPS-X-Au glasses exhibit a distinct absorption peak at 550 nm, corresponding to the characteristic localized surface plasmon resonance (LSPR) of GNIs. As the SiO$_2$ concentration in the NAPS-X glasses increases, the LSPR peak broadens. This indicates that the GNIs formed on NAPS-X glasses lose the ability to absorb light efficiently in a specific wavelength regions and generates the hot electrons. Typically, a sharp LSPR peak is associated with the collective oscillations of free electrons in small size metal nanoparticles. However, as the size of the metal nanoparticles increases, so does their surface area, leading to enhanced damping of these oscillations due to increased electron-electron interactions, electron-phonon interactions, and surface scattering. This damping causes the LSPR peak to broaden and reduces the ability to absorb the light at specific wavelengths. Additionally, as the immersion fraction of the GNIs on the surface of the NAPS-X glasses decreases with increasing SiO$_2$ content, the effective illuminated surface area increases, further contributing to the damping of electron oscillations. In the case of SiO$_2$-Au glass, the absorption peak is very broad, indicating large size and weak immersion of GNIs. Therefore, the observed broadening of the LSPR peak with increasing SiO$_2$ concentration in NAPS-X glasses is attributed to the increase in the size of the GNIs.

**S6: Surface chemical analysis of gold nano-islands through STEM-in-SEM**

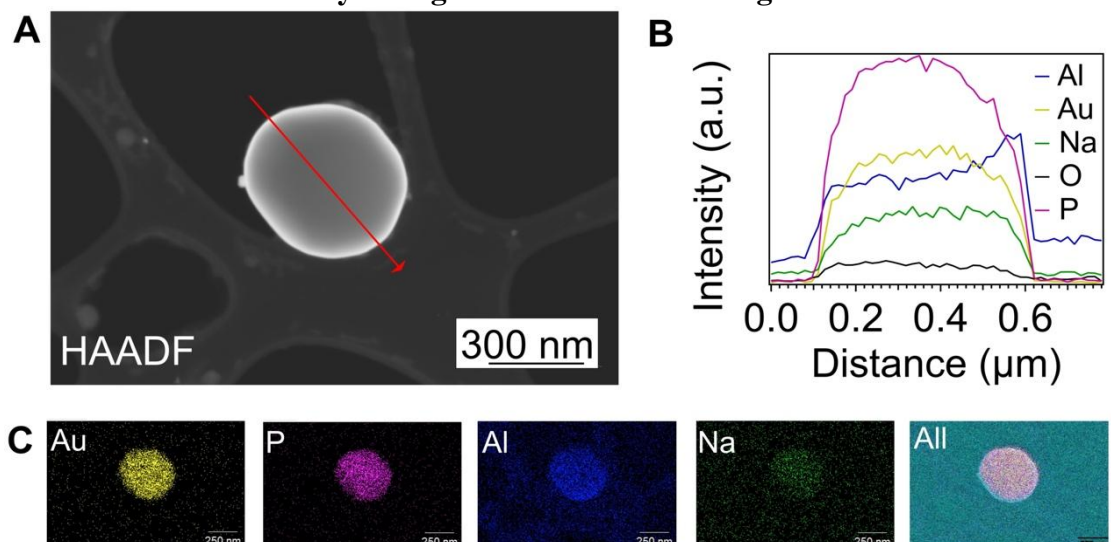

**Figure S4:** (A) HAADF-STEM images (B) Elemental line profiles and (C) Elemental mappings for gold nano-islands (GNIs) extracted from NAPS-30-Au.

**S7: Mechanical stability of gold nano-islands**

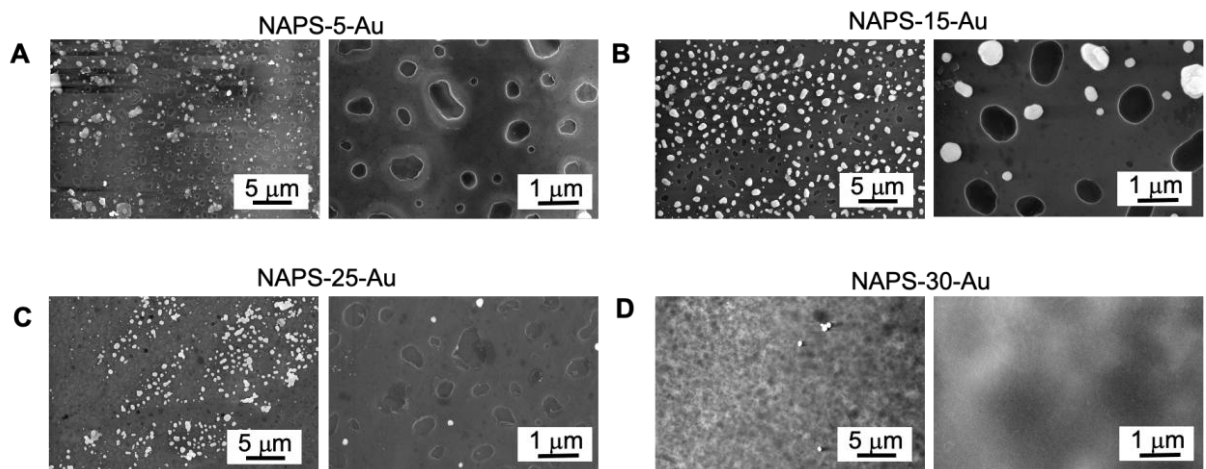

**Figure S5:** FESEM of NAPS-X-Au glasses after extracting the HGNIs through high power ultrasonication.

As an initial step, we tested the stability of the GNIs on NAPS-X-Au glasses using ultrasonication techniques. FESEM and AFM images for NAPS-X-Au glasses after ultrasonication are shown in Figures S4 and S5, respectively. The images reveal the presence of shallow cavities where GNIs were extracted, confirming that the GNIs are deeply integrated into the NAPS-X glasses (X = 5, 15, and 25). The persistence of GNIs in NAPS-15-Au even after high-power ultrasonication highlights their ultrastability. The formation of shallow dips in NAPS-X-Au glasses (X = 5, 15, and 25) undoubtedly confirms that the GNIs are etching the surfaces of the NAPS-X glasses. In contrast, FESEM and AFM image analysis for NAPS-30-Au after GNI extraction shows the absence of GNIs, indicating that the GNIs formed on NAPS-30 are poorly connected to the glass surface. The depth of the cavities decreases with increasing the SiO$_2$ concentration (Figure S5). These observations suggest that the GNIs are chemically interacting with the NAPS-X glass, particularly with phosphorus, as the EDS line profile shows a higher dominance of phosphorus compared to other glass elements. Furthermore, the P$_2$O$_5$ content is higher in NAPS-5 and NAPS-15 glasses compared to NAPS-25 and NAPS-30 glasses, indicating that the presence of P$_2$O$_5$ plays a critical role in enhancing the stability of the GNIs. GNIs are highly stable on NAPS-05-Au glass, however, NAPS-05 glass exhibit the significant shape deformation. The as formed GNIs diffused moderately and exhibiting stability along with the dignified LSPR on NAPS-15 glass, indicating a suitable substrate for the ultrastability of GNIs.

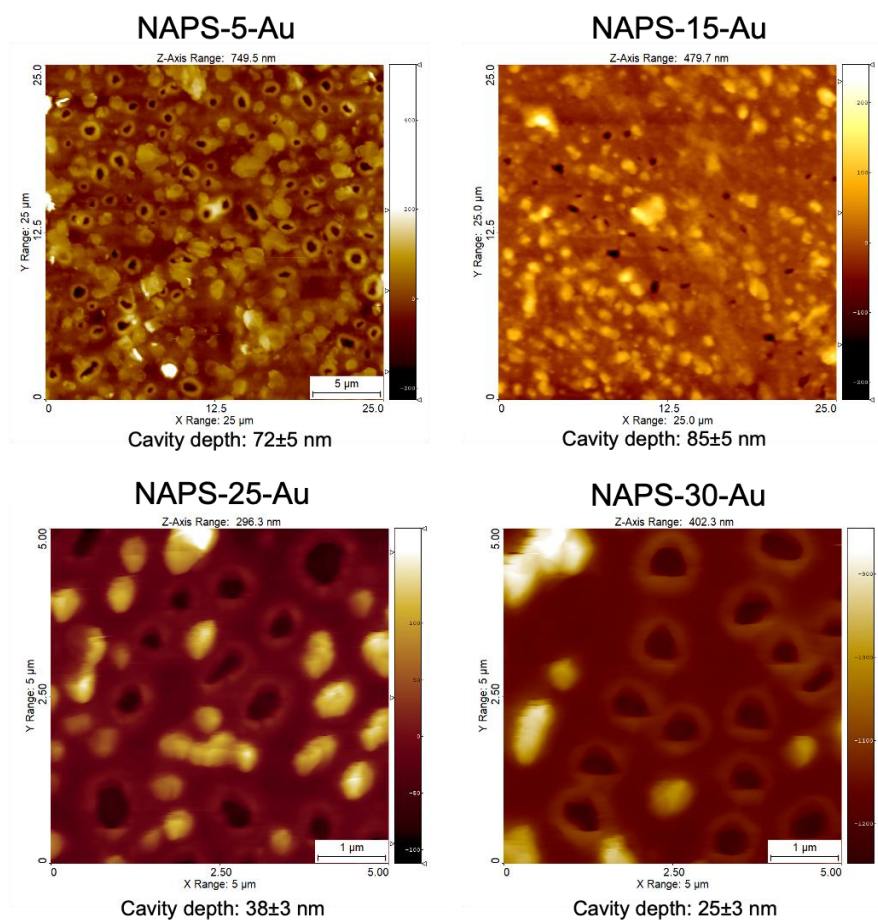

**Figure S6:** (A) 2D AFM images of NAPS-X-Au glasses after extracting the HGNIs through high power ultrasonication.

**S8: X-ray photoelectron spectroscopy**

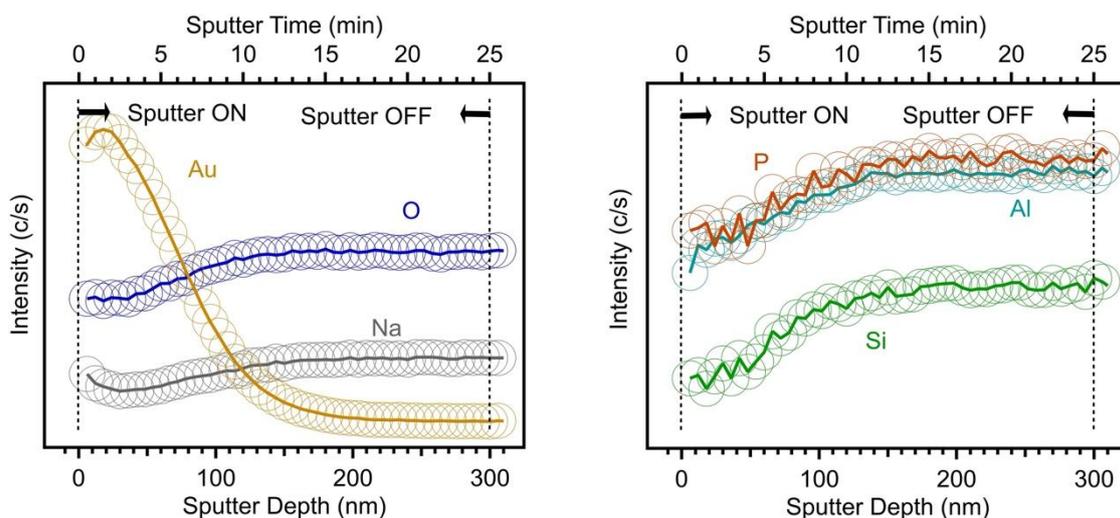

**Figure S7:** Variations in the integrated area under the XPS high-resolution peaks for Au, Na, O, P, Si, and Al as a function of sputtering depth for NAPS-30-Au.

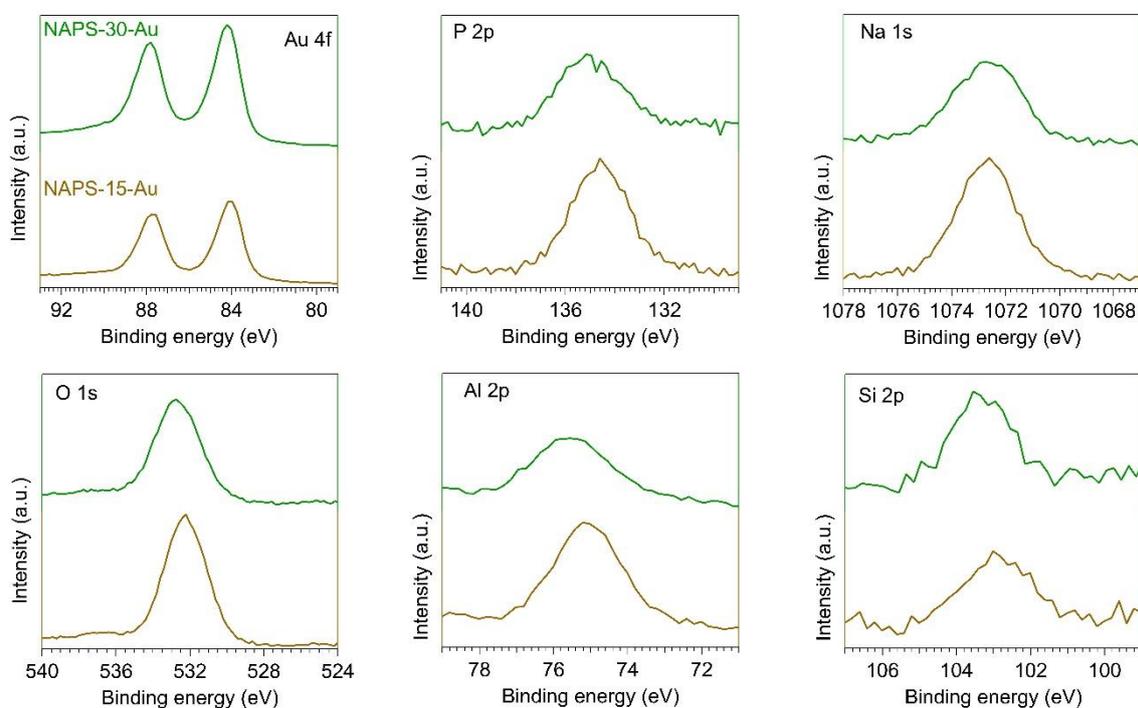

**Figure S8:** High resolution XPS spectra of Au 4f, P 2p, Na 1s, O 1s, Al 2p and Si 2p for NAPS-15-Au glass and NAPS-30-Au glasses.

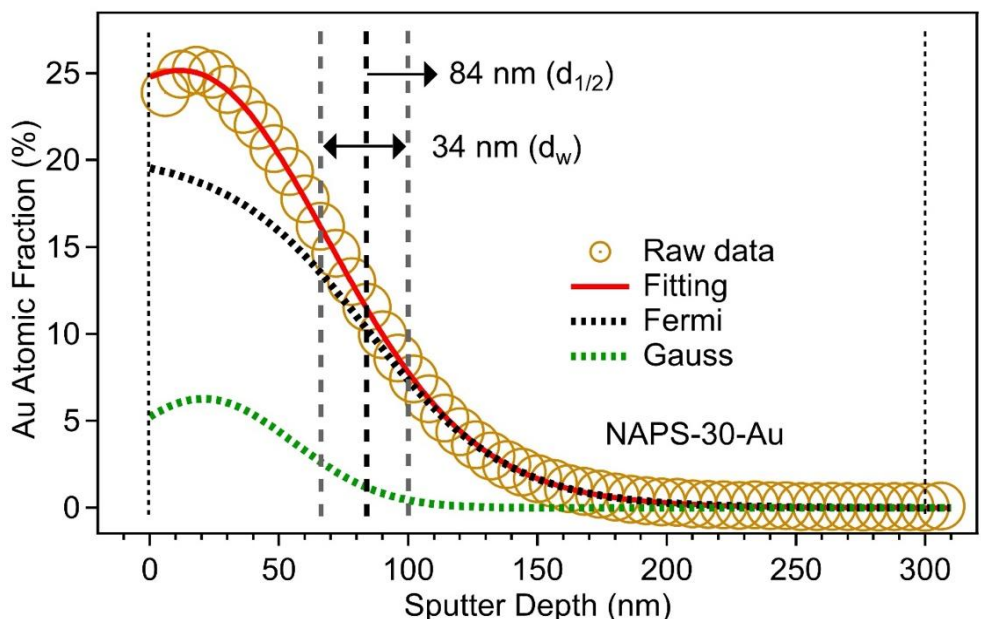

**Figure S9:** Changes in the atomic fraction of Au with sputtering depth, fitted using a convolution of Fermi and Gauss functions for NAPS-30-Au. $d_{1/2}$ is the sputter depth at half the signal intensity maxima, and $d_w$ is the width of the interface between the glass and GNIs.

**Table S1:** Au 4f XPS-depth profile fitting parameters.

|  | NAPS-15-Au | NAPS-30-Au |
| --- | --- | --- |
| $d_{1/2}$ (nm) | 123 | 84 |
| $d_w$ (nm) | 68 | 34 |
| $x$ (nm$^{-1}$) | 16.6 | 27.1 |

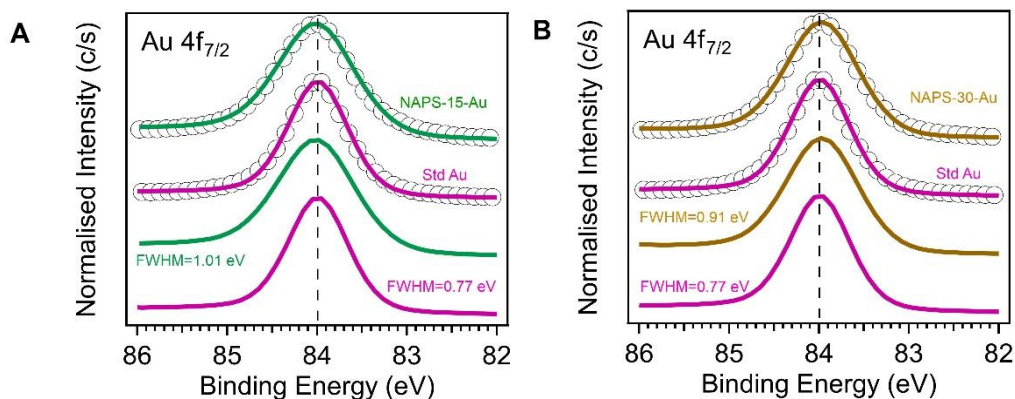

**Figure S10:** Comparision of FWHM of (A) NAPS-15-Au and (B) NAPS-30-Au Au $4f_{7/2}$ peak with the Standard Au $4f_{7/2}$ peak. For comparison purpose and to estimate full widths at half maximum (FWHMs), the marginal shifts are adjusted to align with the standard Au.

     Measured full width at half maximum (FWHM) values for Au $4f_{7/2}$ peak are as follows: 1.01 eV for NAPS-15-Au, 0.91 eV for NAPS-30-Au, versus 0.77 eV for standard Au. The broadening could be a signature of inhomogeneous potential distribution within the probed area as the electrostatic potential is most likely uniform within the same. However, this broadening cannot be surface related, as the surface termination is identical.

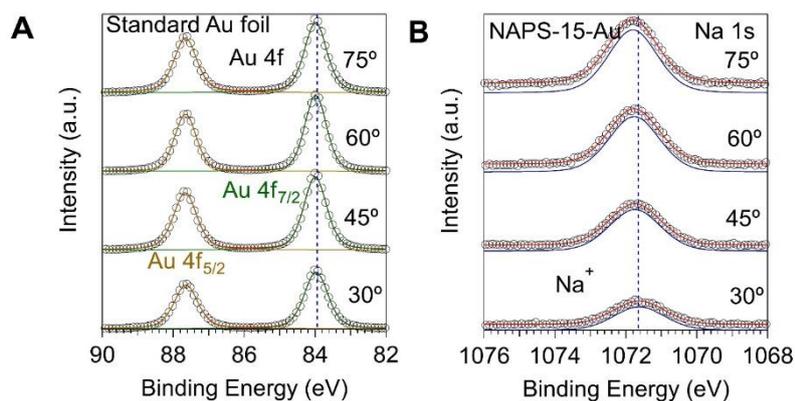

**Figure S11:** Angle dependent XPS peak fitting for (A) standard gold foil Au 4f and (D) NAPS-15-Au Na 1s.

**Table S2:** Angle dependent XPS Fitted Peak parameters

| Element | Spectral Region | Electron Escape Depth w.r.t Analyzer Angle (Å) | | | | Peak FWHM w.r.t Analyzer Angle (eV) | | | |
|---|---|---|---|---|---|---|---|---|---|
| | | 30° | 45° | 60° | 75° | 30° | 45° | 60° | 75° |
| Standard Au | Au $4f_{7/2}$ | 70.28 | 99.38 | 121.72 | 135.76 | 0.79 | 0.76 | 0.72 | 0.74 |
| | Au $4f_{5/2}$ | | | | | 0.80 | 0.77 | 0.72 | 0.75 |
| Au | Au $4f_{7/2}$ | 70.28 | 99.38 | 121.72 | 135.76 | 0.85 | 0.82 | 0.82 | 0.93 |
| | Au $4f_{5/2}$ | | | | | 0.85 | 0.83 | 0.82 | 0.93 |
| P($PO_4^{3-}$) | P $2p_{3/2}$ | 31.83 | 45.01 | 55.13 | 61.49 | 1.54 | 1.52 | 1.48 | 1.46 |
| | P $2p_{1/2}$ | | | | | 1.55 | 1.52 | 1.48 | 1.46 |
| Na($Na^+$) | Na 1s | 47.28 | 66.86 | 81.89 | 91.34 | 1.68 | 1.71 | 1.68 | 1.66 |

**S9: Transmission Electron Microscopy**

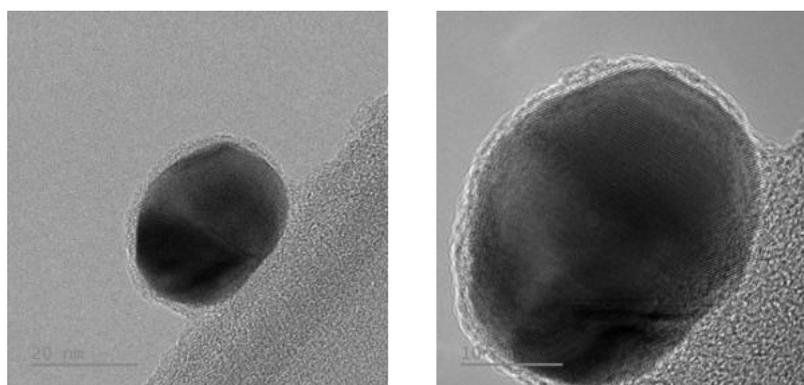

**Figure S12:** Low resolution transmission electron microscopy images of Figure 2D.

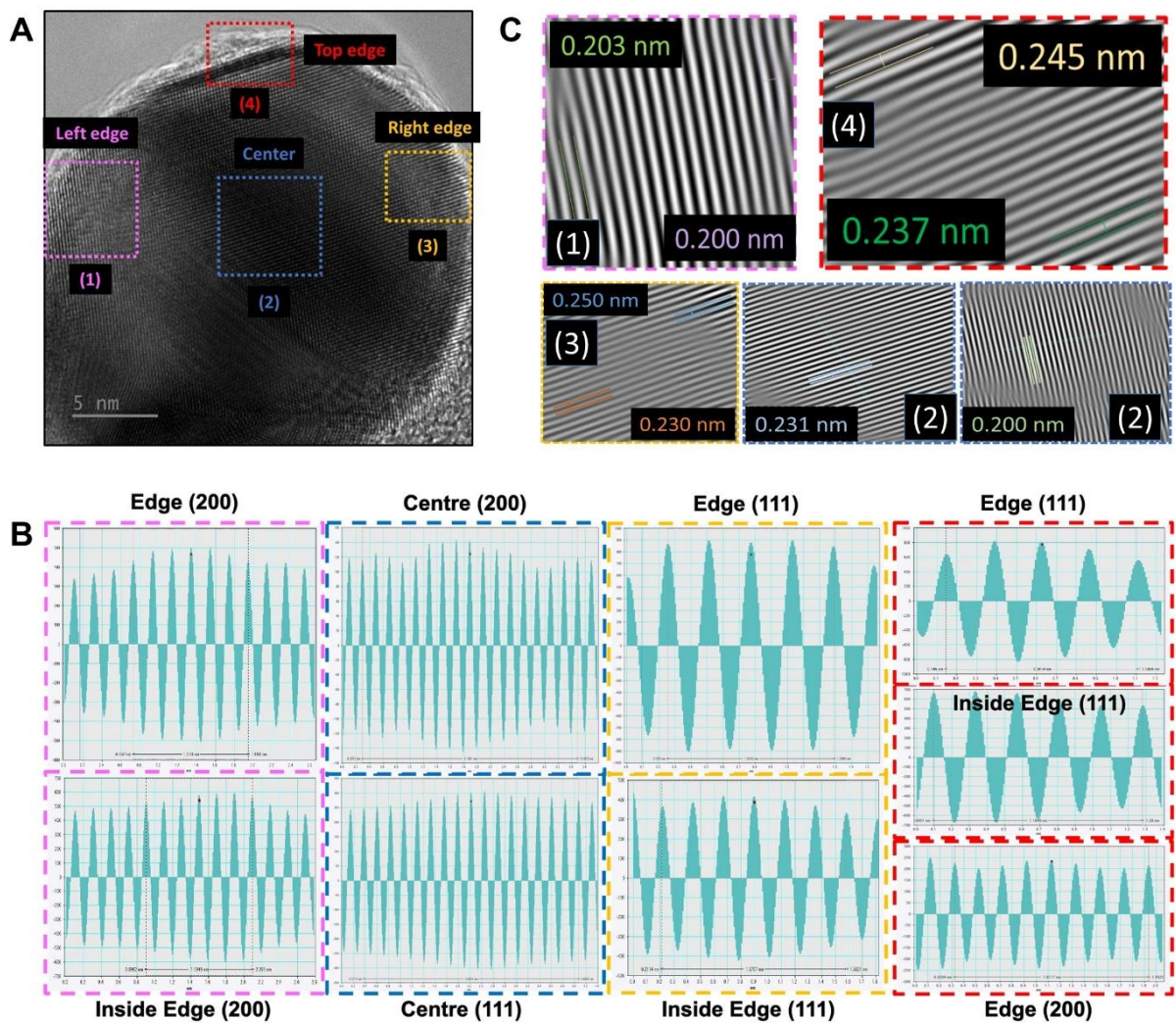

**Figure S13:** (A) HRTEM image of GNI extracted from NAPS-15-Au. (B) Inverse Fast Fourier Transformed images (IFFT) image for the regions highlighted in (A). (C) live IFFT images for the regions highlighted in (A).

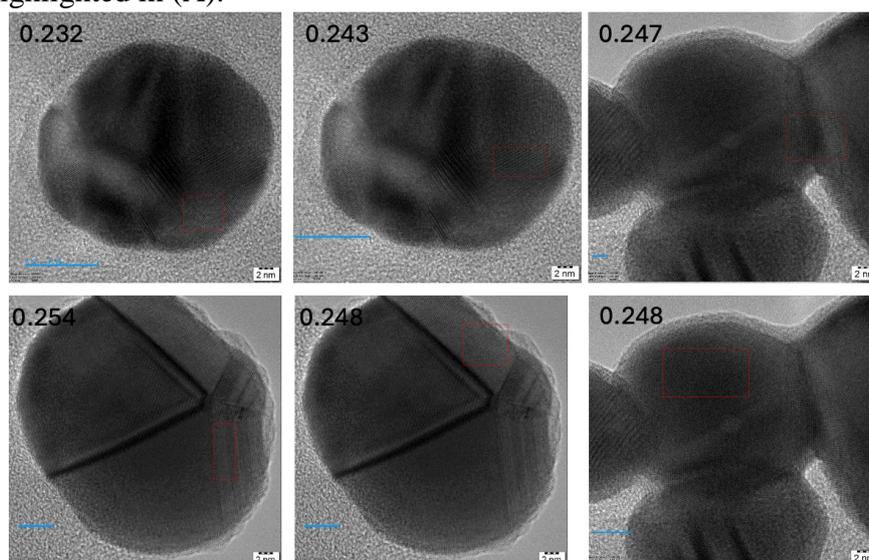

**Figure S14:** HRTEM images for GNIs extracted from NAPS-15-Au.

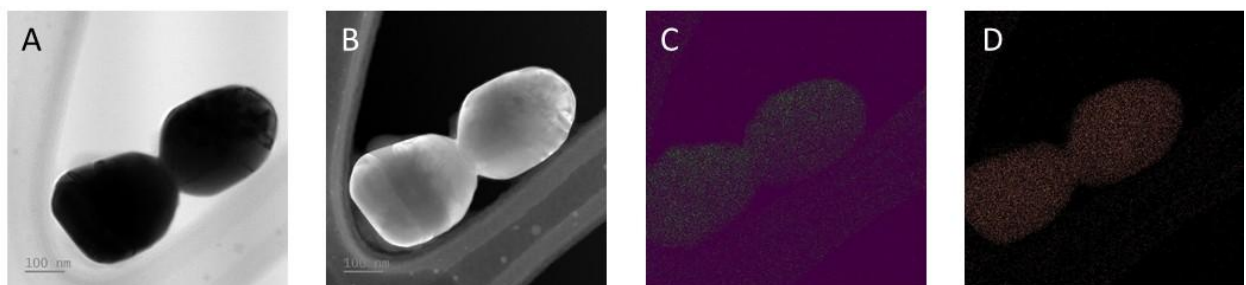

**Figure S15:** (a) STEM; (b) Dark field STEM image for HGNI presented in Figure 2F. (C) oxygen and (d) silicon elemental mapping in HAADF for HGNI presented in Figure 2F.

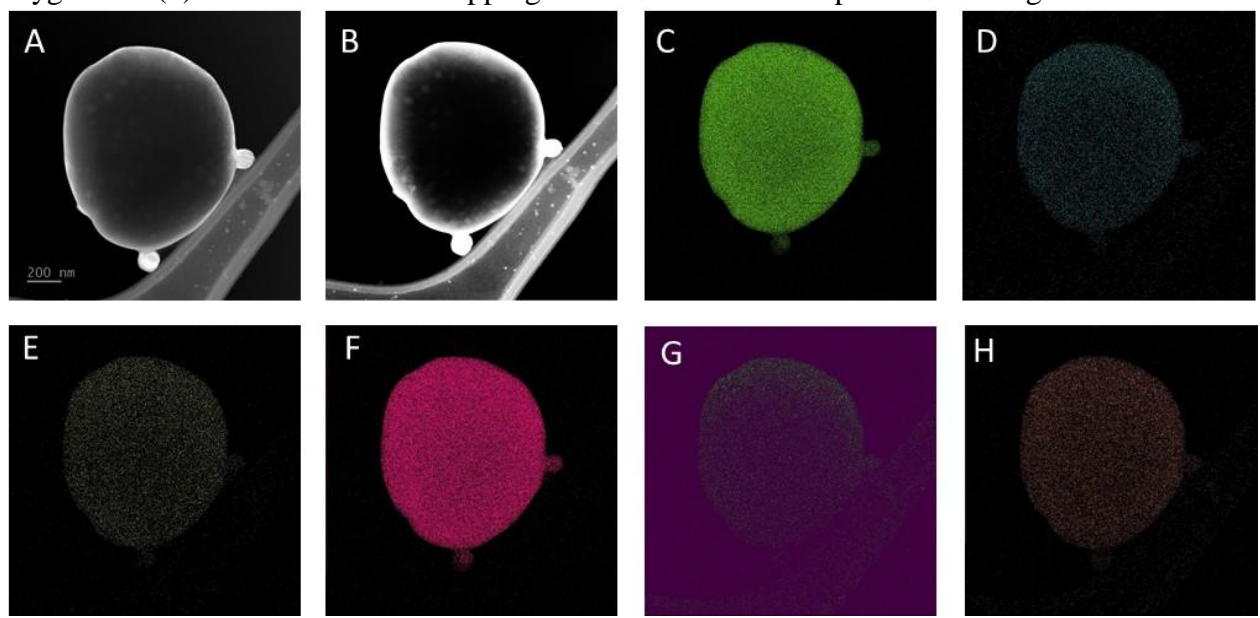

**Figure S16:** (A) HAADF-STEM image; (B) Dark field HAADF-STEM image and elemental mapping for (C) Au, (D) Na, (E) Al, (F) P, (G) O and (H) Si for HGNI.

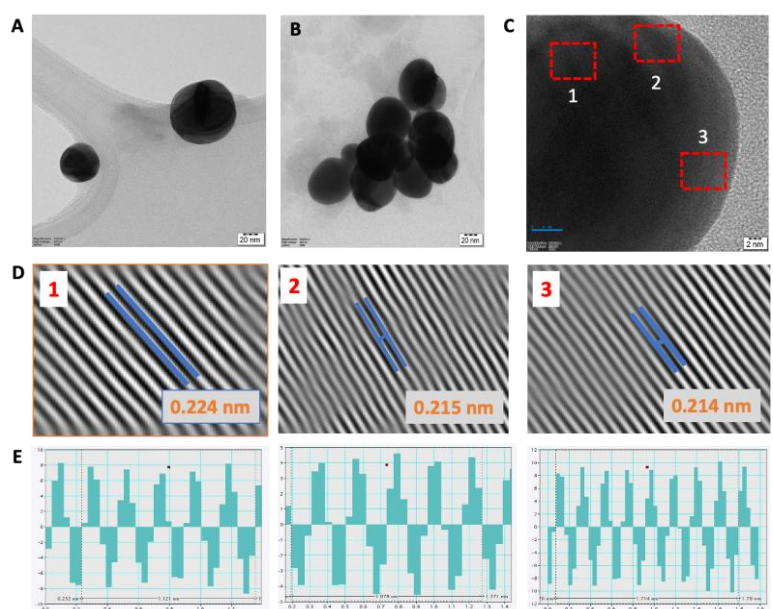

**Figure S17:** (A) Bright field (BF) TEM image of particle extracted from standard Silicon dioxide substrate, (B) Clusters of extracted particles from standard substrate, (C) High

resolution lattice imaging of extracted particle displaying lattice fringes, (D) live FFT of regions marked in (C) and (E) IFFT line profile regions marked in C.

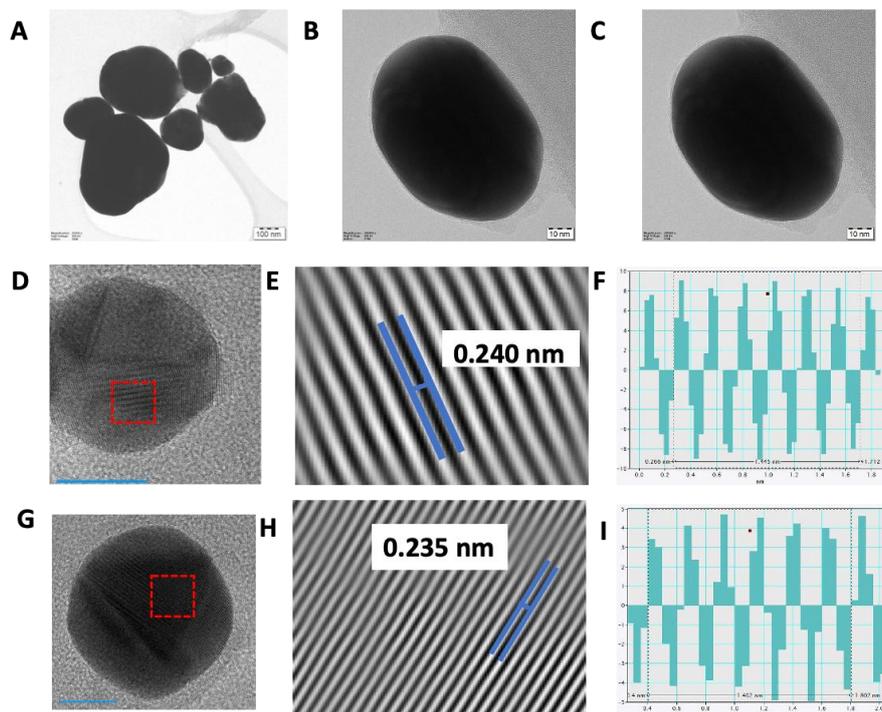

**Figure S18:** (A-C) Bright field (BF) TEM image of particle extracted from NAPS-30-Au substrate, (D, G) High resolution lattice image of extracted particle with visible fringes, (F, I) line profile of lattice across small region indicating average value of 0.240 nm as interplanar spacing and 0.235 nm respectively and (E, H) live IFFT images for the highlighted region of the figures (D, G).

Over a large number of (111) lattice spacing measurements from various GNIs, an average increase of *d* spacing is observed (Figure S14). Figure S17 shows TEM images of GNIs extracted from $SiO_2$-Au substrates. Panels A–B and F display the TEM images of the particles, while panel C shows the lattice image of one such particle for d-spacing analysis. Two regions were analyzed using ImageJ, and their d-spacing profiles (avg of 0.22 and 0.23 nm) are presented in E based on FFT analysis from panel D. The data demonstrates that there is only a nominal change in the d-spacing, and no multiple interfaces are present.

TEM images for GNIs extracted from NAPS-30-Au are presented in Figure S18. Panels A–D illustrate their morphology, revealing that the particles are spherical and well-extracted. The electron-transparent regions were further analyzed using FFT, and the d-spacing profiles for particles shown in panels E and F are depicted in panels F and H, respectively. The average d-spacing values were found to be 0.241 nm and 0.235 nm.

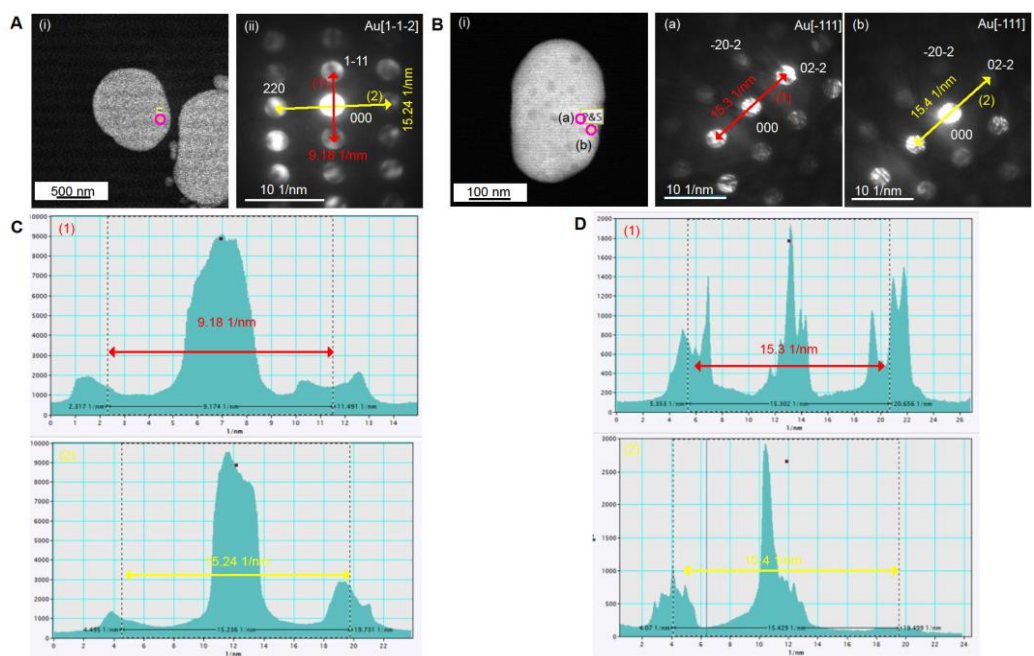

**Figure S19:** (A) (i) The high magnification HAADF image of gold island nanoparticle extracted from NAPS-15-Au. (ii) The nanobeam diffraction pattern (NBD) acquired along the Au[1-1-2] zone axis from the "O" mark region of Au nanoparticle shown in (i). (B) (i) The high magnification HAADF image of another gold island nanoparticle extracted from NAPS-15-Au. (ii) The nanobeam diffraction pattern (NBD) acquired along the Au[-111] zone axis from the "O" mark region (a) of Au nanoparticle shown in (i). (iii) The nanobeam diffraction pattern (NBD) acquired along the Au[-111] zone axis from the "O" mark region (b) of Au nanoparticle shown in (i). (C) Histogram analysis corresponding to Figure A(ii). (D) Histogram analysis corresponding to Figure B(ii) and B(iii).

*S10: Molecular dynamics simulations and Thermal properties of NAPS-X glasses*

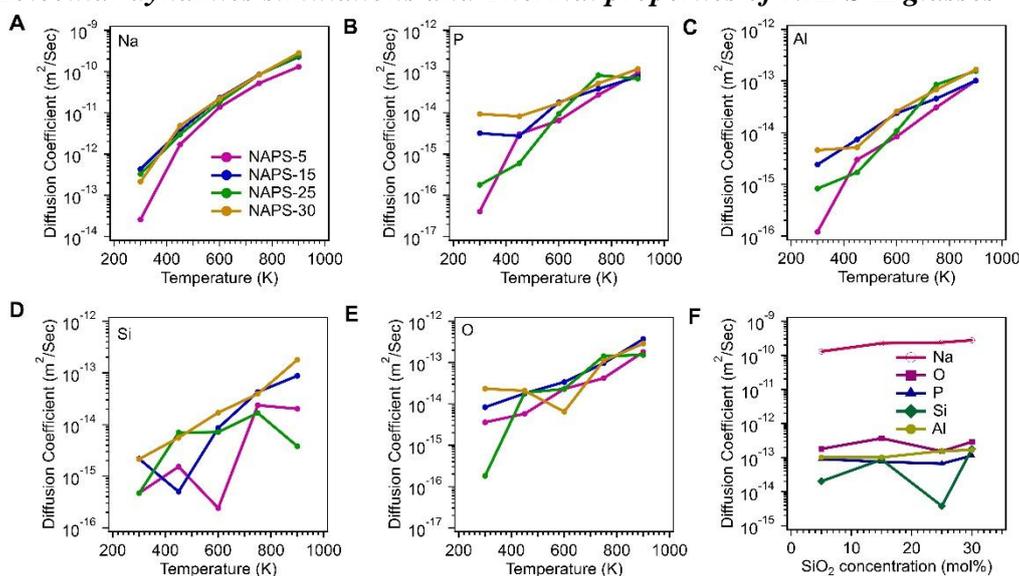

*Figure S20:* Variation in diffusion coefficient of (A) Na (B) P (C) Al (D) Si and (E) O with temperature in NAPS-X glasses. (F) Variation in diffusion coefficient of elements exist in NAPS-X glass at 900 K.

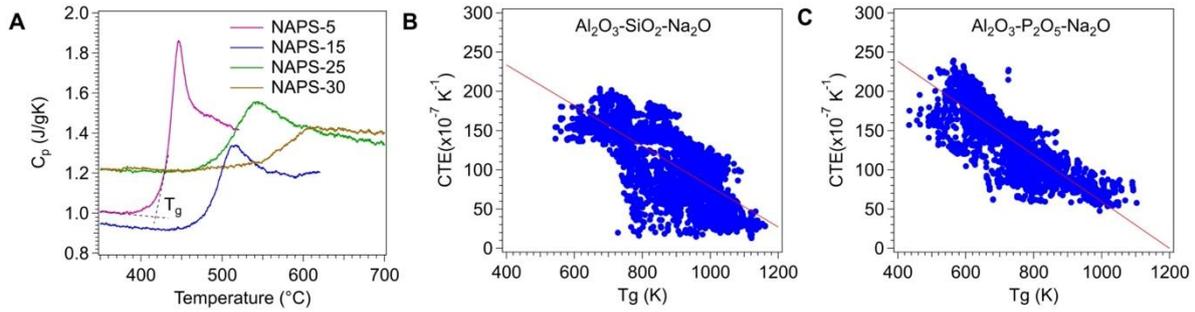

**Figure S21:** (A) Specific heat measurements for NAPS-X glass. Correlation between the coefficent of thermal expansion (CTE) and the glass transition temperature ($T_g$) for (B) $Al_2O_3$-$SiO_2$-$Na_2O$ and (C) $Al_2O_3$-$Na_2O$-$P_2O_5$ glasses (Source: Pyggi Software).

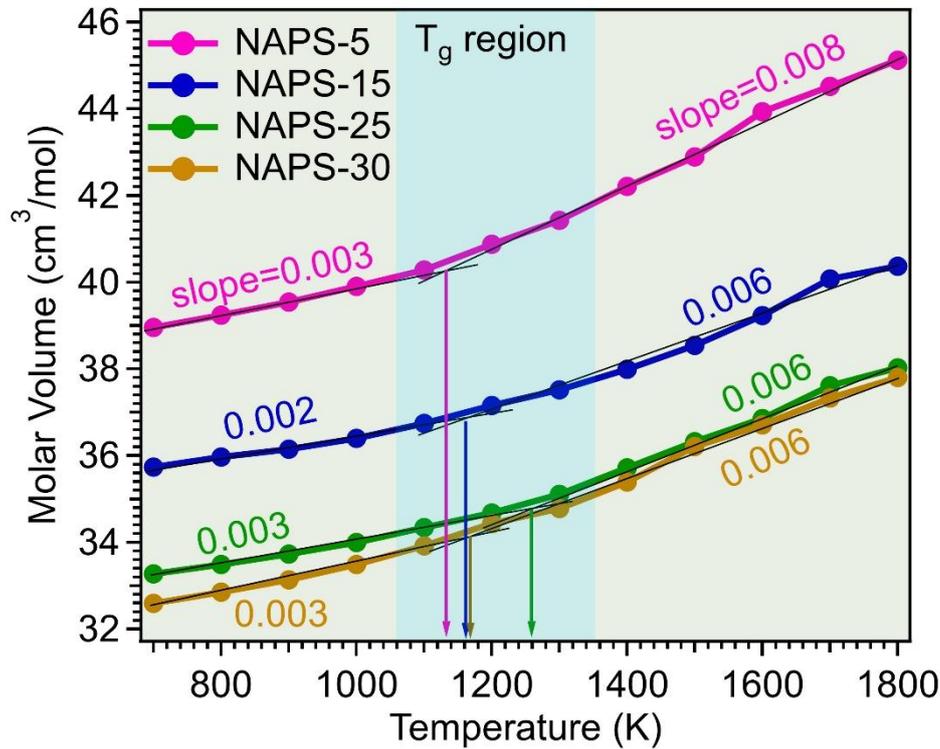

*Figure S22:* Variation in molar volume with temperature (extracted from glasses synthesized using MD simulations).

   The viscoelastic behavior of the support material, as influenced by temperature, is essential for the embedding of MNPs (*13*). The $T_g$ of NAPS-X glasses increases with increasing $SiO_2$ substitution, however, the $T_g$ decreases for NAPS-30. This variation in $T_g$ values is likely due to the formation of Si-O-Al and Si-O-Si bonds, which strengthen the cross-linking network, at the expense of P-O-Al bonds. Moreover, the dissociation energy of the Si-O bond (~798 kJ/mol) is higher than that of the P-O bond (~597 kJ/mol). The decrease in $T_g$ for NAPS-30 should be attributed to the distribution of structural units, especially the formation isolated phosphorus structural units from the aluminosilicate network structure. It is well established that an increase in $T_g$ correlates with a decrease in the thermal expansion coefficient of glass (Figure S21B and S21B) (*14*). Additionally, recent findings by Peter et al (*14*) shows that the volume thermal expansion coefficient of glasses ($α_g$) can triple when crossing the $T_g$ during heating (Figure 3B inset). This indicates that the configurational changes, which include structural rearrangement through bond modifications (Figures S3A and S3B), play a key role in determining the volume thermal expansion coefficient of liquid glass ($α_l$) above **$T_g$**. When the temperature exceeds the glass transition temperature ($T_g$), the glass network expands isotropically (Figure S22), resulting in a decrease in viscosity and surface tension. The $T_g$ values for NAPS-X glasses (Figure 3B) are below the thermal dewetting temperature of 550 °C (823 K), indicating substantial softening at this temperature. The increased concentration of $SiO_2$ has significantly reduced the $T_g$ and simultaniouslt the $V_m$ for NAPS-30, likely reducing GNIs diffusion into the network structure of glass during the thermal dewtting. It has been well documented that the molar volume for the aluminophosphate glasses is much higher than that of aluminosilicate glasses (*15*, *16*).

It is noteworthy that solid-state dewetting typically involves hole formation, growth, and ligament breakup at temperatures well below the film melting point. While the melting temperature of bulk gold is 1337 K, that of ~5 nm Au nanoparticles is approximately 1150 K (*17*). Given that glass materials are **poor thermal conductors** (thermal conductivity ~0.5 W/mK), it is expected that GNIs embedded within the NAPS-X glass network would experience **localized heating** due to **thermal contact resistance** at the GNI/glass interface (*18*). This localized heating could induce **interface melting at temperatures lower than expected**, facilitating the formation of a smooth interface between GNIs and the surrounding glass. A smooth interface would, in turn, enhance **interdiffusion of ions and atoms between the glass substrate and GNIs**, resembling processes such as the **$Ag^+$–$Na^+$ ion exchange** (*19*). The reduction in diffusion of GNIs in to the glasses raises the melting temperature of the GNIs due to the poor heat reflections and limits chemical interactions between the glass and metal. Despite of high diffusion coefficient for Na, it is noted that the diffusion coefficient of all the elements exist in NAPS-X glasses are independent on the chemical composition. These findings indicate that the observed decrease in $V_m$ and increase in $T_g$ with increasing $SiO_2$ concentration suggests a key role in controlling the extent of GNI diffusion into the glass matrix.

It has been reported that the dissociation energy for Si from Si-O-Al is higher than that for Si-O-Si bonds. MD simulations for NAPS-X glasses reveal that $SiO_4$ tetrahedral units are predominantly connected to $AlO_4$ tetrahedral units. Therefore, the dissociation of Si and Al from Si-O-Al bonds in the NAPS-X glasses due to the presence of Au can be diminishes. The higher electronegativity of P (2.2) compared to Si (1.9) suggests a greater affinity of P for Au. FESEM-EDS elemental line profile analysis confirms the enrichment of each island with P, indicating the interaction of Au more effectively with P. It is, therefore, confirmed that the Na due to the high mobility and P due to high electronegativity are majorly interacting with the GNIs and leading to the formation of multi-element GNIs (MEGNIs).

**S11: AFM, SEM, Ex-situ and in-situ XRD analysis of Au film coated NAPS-15 glass**

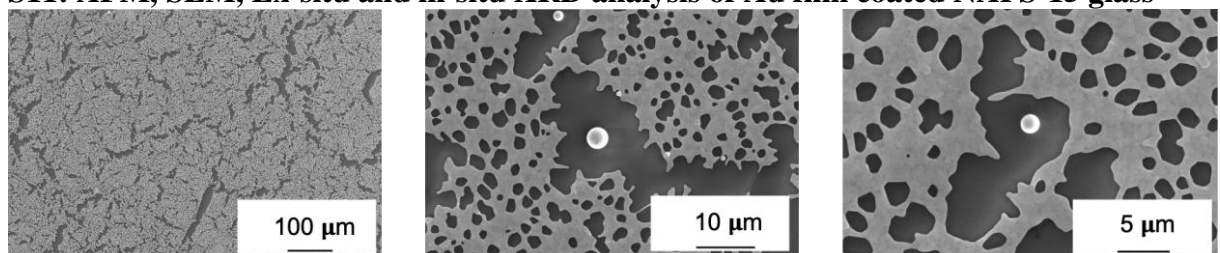

**Figure S23:** SEM images of thermally deposition gold film on NAPS-15 glass substrate followed by heat treatment at 550 °C for 15 min in ambient conditions.

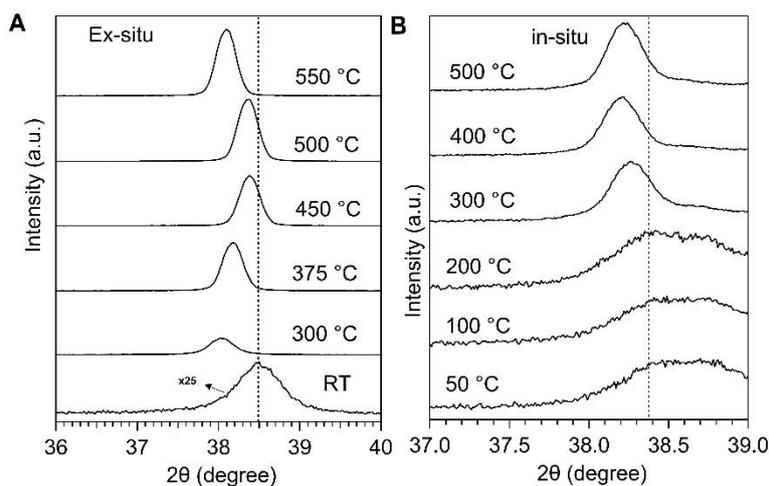

**Figure S24:** (A) Ex-situ XRD spectra for Gold films depositied on different substrated and heat treated at 300, 375, 450, 500 and 550 C for 15 min. (B) In-situ XRD spectra for Gold films depositied on NAPS-15 substrate.

Our observations indicate that the Au film deposited on NAPS-15 glass via thermal evaporation remains continuous and does not exhibit island morphology (Figure S23), highlighting the influence of the deposition technique on film morphology. To study the dewetting behaviour of a 50 nm thick Au film deposited by DC sputtering, we conducted ex-situ AFM and FE-SEM analyses on NAPS-15 glass substrates. The 50 nm Au film was deposited on various NAPS-15 substrates and subjected to heat treatments at 300 °C, 375 °C, 450 °C, 500 °C, and 550 °C for 15 min each in an air atmosphere, followed by FE-SEM (Figure 3D(bottom)) and AFM (Figure 3D(bottom)) analysis. FE-SEM images (Figure 3D(bottom)) show that film rupture and perforation began heterogeneously in multiple areas at the relatively low temperature of 300 °C, suggesting that the 50 nm Au film started melting below 300 °C. While interconnected islands remained, complete dewetting of the Au film was observed at 375 °C. As the temperature increased from 375 °C to 550 °C, the number of interconnected islands decreased, leading to the formation of numerous spherical islands. We also deposited 50 nm-thick Au films on unique NAPS-15 substrate and performed sequential thermal treatments at 300 °C, 375 °C, 450 °C, 500 °C, and 550 °C for 15 minutes in air. AFM analysis (Figure 3D(top)) reveal GNI mean height (Figure 3E) initially increases from 300 °C to 450 °C, followed by pronounced reduction at higher temperatures. The comparison of morphological images of GNIs on different glass substrates with those on the same substrates at identical heat treatment temperatures reveals significant variations (Figure 3D). These differences may be due to the use of furnaces located in different geological regions. However, the mean height of the GNIs (Figure 3E) decreases sharply after exceeding $T_g$, highlighting the strong influence of $T_g$.

Ex-situ XRD spectra for NAPS-15-Au at different temperatures are shown in Figure S24A. The XRD spectra reveal that the 2θ position corresponding to the (111) crystalline plane of FCC Au shifts towards lower angles as the temperature increases. This change in XRD peak position could be attributed to combined effects of thermal dewetting, lattice strain, and interdiffusion between the gold nano-islands and the glass substrate. Generally, the shift to lower 2θ indicates an increase in d-spacing. During the formation of nano-islands due to thermal dewetting of the Au film, Na atoms, due to their out-diffusion behavior, and P atoms may occupy the grain boundaries or the interstitial sites of FCC Au. Given the atomic radii of Au (144 pm), Na (186 pm), and P (110 pm), phosphorus is smaller than Au, while sodium is larger. The increase in d-spacing could be attributed to the strain in FCC Au due to the Na occupation in FCC Au. However, it remains challenging to make definitive comments about the specific atomic sites of Na and P. In-situ high-temperature XRD measurements (Figure S24B) confirm the decrease in 2θ position corresponding to the (111) crystalline plane. It is important to note that the reaction between gold and NAPS glass occurred under air atmospheric conditions. Therefore, correlating ex-situ and in-situ XRD measurements for this system is challenging, as in-situ XRD measurements were performed under vacuum conditions.

## S12: Density Functional Theory

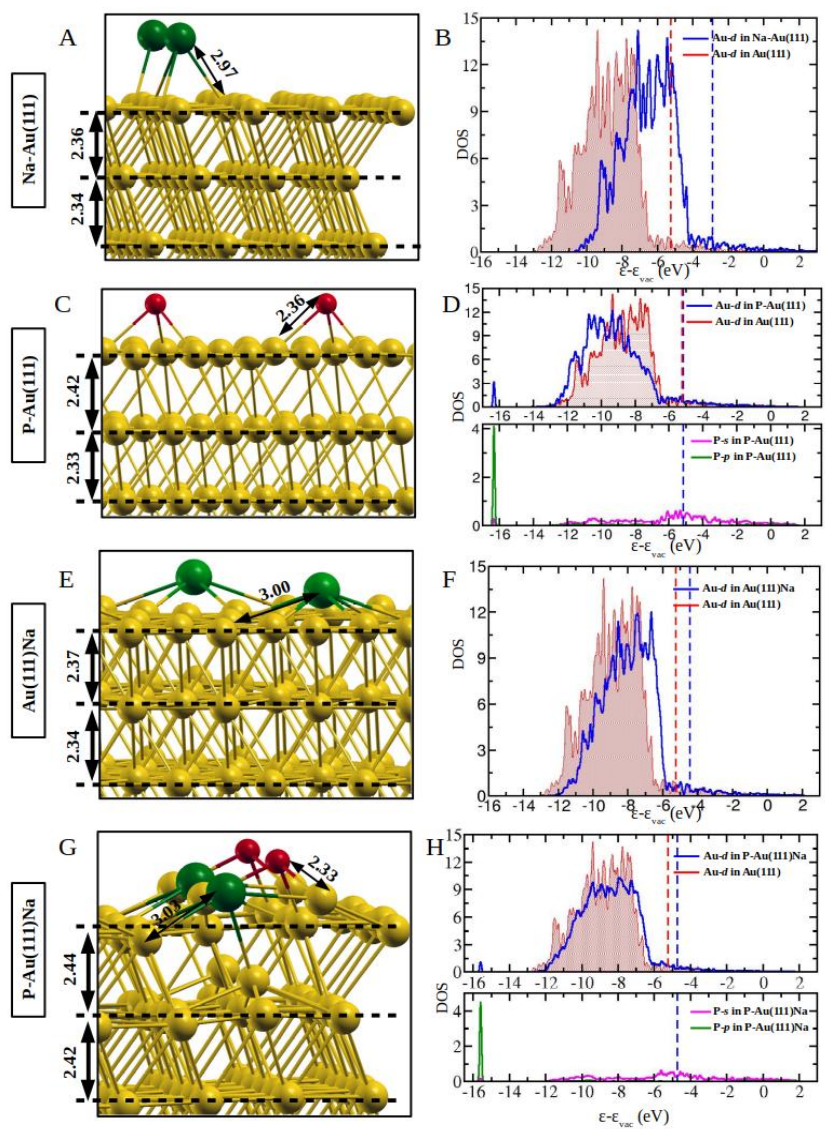

**Figure S25:** Structure and DOS for different doped Au(111) slab models. Distances are noted in Angstrom unit. Energies in DOS are shifted with respect to Fermi energy. The dotted vertical lines in red and blue denote the Fermi energies of clean Au(111) and doped Au(111) slab, respectively.

**Table S3:** Comparison of bondlengths in different slab models studied in this work. For the doped models Au-Au distance is noted for those Au which bind to the dopants. The distances are noted in Angstrom unit.

| Model | Au-Au | Na-Na | Na-P | P-P | Na-Au | P-Au |
|---|---|---|---|---|---|---|
| Au(111) | 2.83 | - | - | - | - | - |
| Na-Au(111) | 2.87 | 8.50 | - | - | 2.97 | - |
| P-Au(111) | 3.04 | - | - | 8.50 | - | 2.36 |
| Au(111)Na | 2.83 | 8.50 | - | - | 3.00 | - |
| NaP-Au(111) | 2.88 ± 0.10 (Na); 3.43 ± 0.20(P) | 8.50 | 4.29 | 8.50 | 3.07 | 2.35±0.02 |
| P-Au(111)Na | 2.79± 0.07 (Na); 3.38 ± 0.20 (P) | 8.50 | 3.59 | 8.50 | 3.04 | 2.34±0.04 |

In Figure S25 we have shown the structure and density of states plots for Na-Au(111), P-Au(111), Au(111)Na and P-Au(111)Na. As noted in the main text, we observe minimal distortions for slabs doped with single glass element. However, for P-Au(111)Na we notice distortions spreading up to the second layer from top. We can see Au atoms moving out-of-plane from top layer as well as from second layer from top in P-Au(111)Na. This results in larger interplanar distances in comparison to those in clean Au(111) or deeper layers in the same doped slab. Again as noted in the main text Na does not participates in chemical interactions with the surface Au and hence we observe just a shift in Au-d of Na-Au(111) and Au(111)Na. However, s and p states of P are seen to hybrise with Au-d in P-Au(111) and P-Au(111)Na which result in a much stronger P-Au bond as compared to the weaker Na-Au bond.

## S13: Transient Absorption Spectroscopy

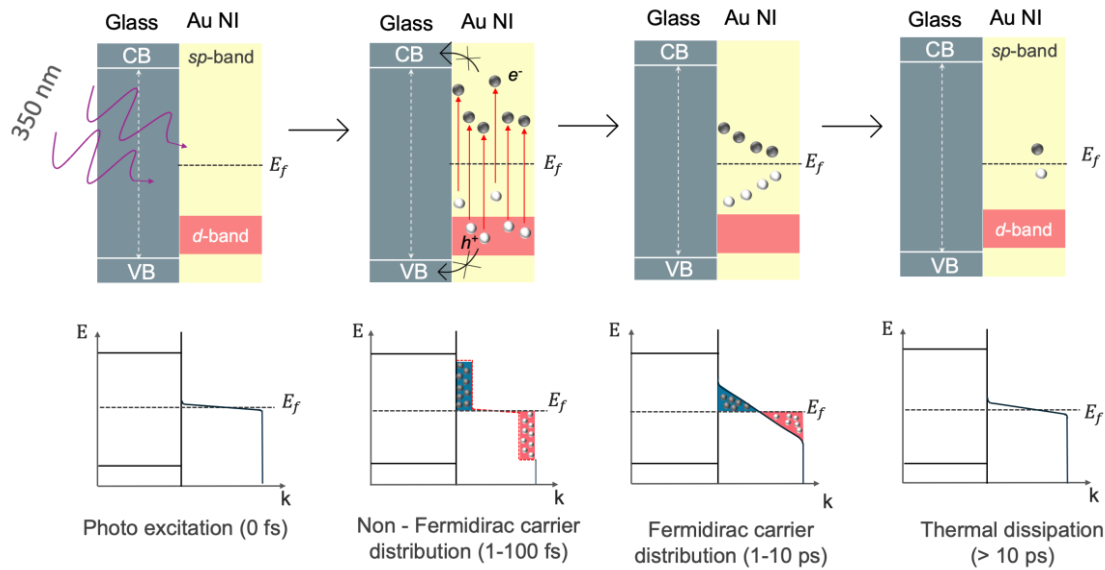

**Scheme S2:** Schematic of the hot electron formation and their relaxation in MEGNIs supported by the glass materials

According to the two temperature model (TTM), the rate of energy exchane between the electrons and phonons are given by the following equations (S1) and (S2):
$$C_e(T_e)(dT_e/dt)=-g(T_e-T_l)\ldots (S4)$$

$$C_l(dT_l/dt)=g(T_e-T_l)\ldots (S5)$$

where $T_e$ and $T_l$ are the electronic and lattice temperatures, $C_l$ is the lattice heat capacity, $C_e(T_e) = \gamma T_e$ is the temperature dependent electronic heat capacity, g is the electronphonon coupling constant and γ is the electron heat coefficient (66 J/m$^3$K$^2$ for gold at room temperature $T_0$). The eq. (S4) and (S5) (*20*), indicate that electron-phonon coupling ($\tau_{e-ph}$) time depends on the initial electronic temperature ($T_e$) and are interconnected via $\tau_{e-ph} = \gamma(T_0+T_e)/g$.

To harness the potential of firmly embedded MEGNIs, it is essential to comprehensively understand the dynamics of hot carriers. The decay of plasmon resonance or the excitation of electrons in MNPs with femtosecond laser pulses generates a non-thermal distribution of hot carriers (electrons and holes) that deviate from the Fermi-Dirac distribution, both above and below the Fermi level. These non-thermal hot carriers rapidly undergo elastic electron-electron (e-e) scattering, leading to their thermalization to the electronic temperature ($T_e$). Consequently, the electron energy distribution conforms to the Fermi-Dirac distribution with an elevated $T_e$. Subsequently, these thermalized hot electrons transfer their energy to the lattice and then to the surrounding environment through electron-phonon (e-ph) and phonon-phonon (ph-ph) coupling, respectively (Scheme S2) (*21*).

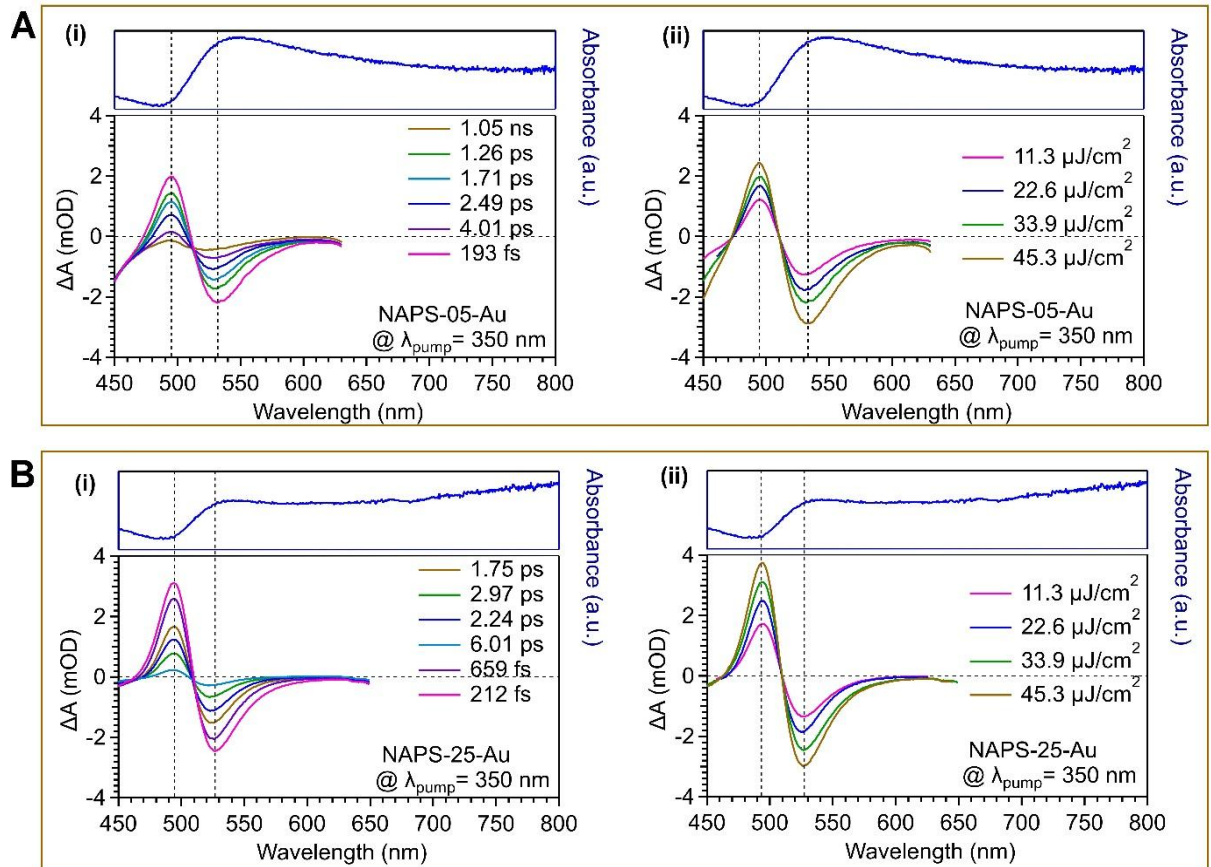

**Figure S26:** Transient absorption spectra (i) at different delay time and at (ii) different pump power for (A) NAPS-05-Au, (B) NAPS-25-Au glasses.

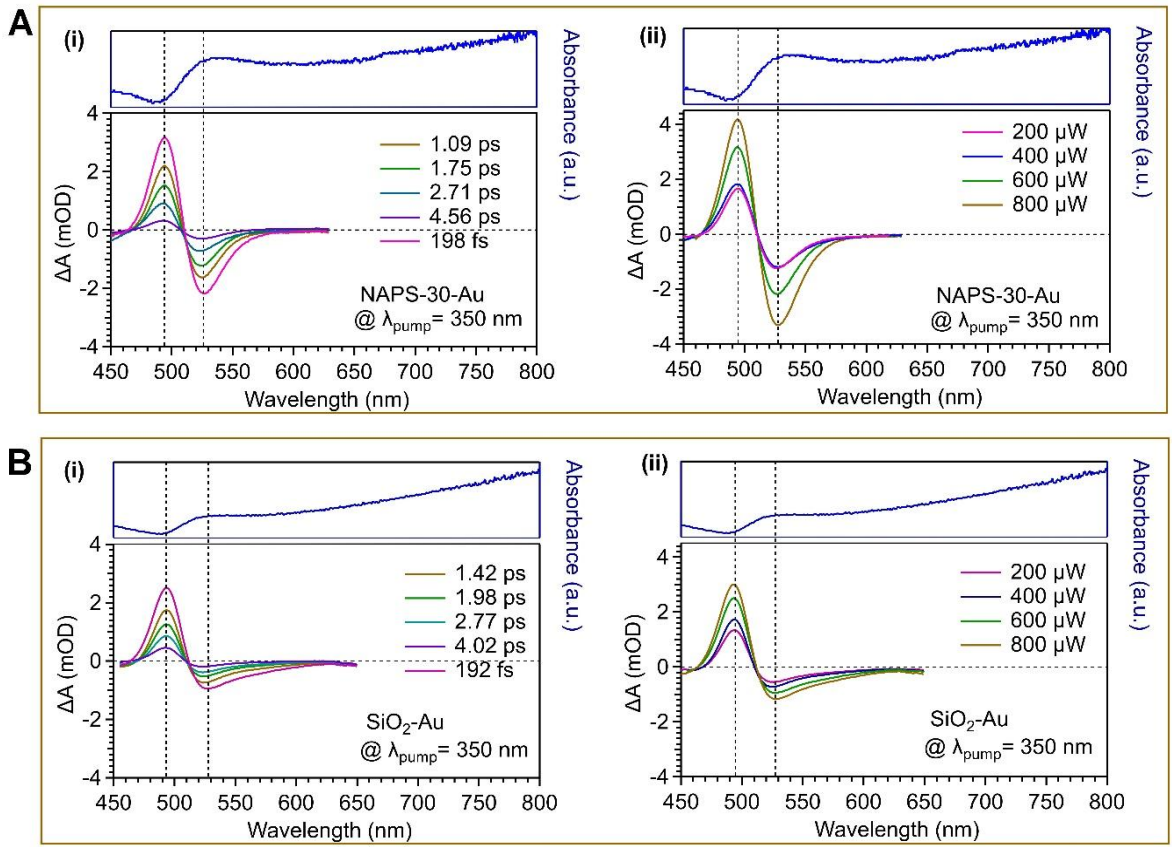

**Figure S27:** Transient absorption spectra (i) at different delay time and (ii) at different pump power for (A) NAPS-30-Au and (B) SiO$_2$-Au glasses.

To assess the decay of the thermalized electron temperature (T$_e$) of MEGNIs through electron-phonon (e-ph) and phonon-phonon (ph-ph) interactions, we recorded transient absorption (TA) spectra for all the glasses at various delay times, ranging from 1 fs to 7 ns, using a 350 nm pump wavelength at different pump powers (200 μW - 800 μW). The TA spectra at different delay times and at varying pump powers with a delay of approximately 200 fs are presented in Figure S26 and S27. For comparison, the TA spectra of SiO$_2$-Au glasses were also recorded and are shown in Figure S27B. Except for differences in the intensity and broadening of the negative ground state bleaching (GSB) peak at 530 nm and the positive excited state absorption (ESA) peak at 490 nm, all the NAPS-X-Au glasses exhibit similar TA spectra across different pump powers and delay times. Highly intense excited state absorption (ESA) typically occurs due to interband electronic transitions (*22*). Additionally, it can result from the absorption of photons by excited electrons, causing their transition to higher energy levels.

The obtained τ$_{e-ph}$ for all the samples at different pump fluences is shown in Figure 5D. τ$_{e-ph}$ increases linearly with increasing pump fluences, indicating the validation of the TTM for the present system. Several studies noted that the thermal conductivity of the environment significantly influences τ$_{e-ph}$; an increase in the thermal conductivity of the environment decreases τ$_{e-ph}$ (*23, 24*). AFM image analysis confirms that the NAPS-X glass environment around the MEGNIs decreases with increasing SiO$_2$ concentrations. Knowing that air is a poorer thermal conductor (0.026 W/mK) than NAPS-X glasses (~0.5 W/mK), the τ$_{e-ph}$ should increase with decreasing the contribution of NAPS-X glass environment. In addition to the thermal conductivity of the substrates, other parameters are also associated with the dynamics observed for MEGNIs embedded on NAPS-X glasses. Further studies are needed to better understand the Plasmon decay and energy transfer mechanism within the MEGNI and between MEGNIs and glass.

At zero pump power, the rise in temperature will be zero. The life time of hot electrons ($\tau^0_{e-ph}$) at zero pump power and temperature-independent g relate via $\tau^0_{e-ph} = \gamma T_0/g$. We extracted $\tau^0_{e-ph}$ for all the samples by extrapolating the linear plots to zero absorbed energy density to evaluate the g values. The obtained $\tau^0_{e-ph}$ for all the NAPS-X glasses are shown in the Figure 5G. The g value for MEGNIs decreases with increasing $SiO_2$ concertation in NAPS-X-Au glasses up to 25 mol% and then increases for NAPS-30-Au glass. This concludes that the chemical composition of MEGNIs have influences the variations in $\tau_{e-ph}$, which, however, could be attributed the variations in the density of available electronic states (DOS) near the Fermi level. It has been reported that, irrespective of the size of polycrystalline Au, grain boundaries significantly decrease the $\tau_{e-ph}$ (*25*). We attribute the difference in $\tau_{e-ph}$ value to the existence of multiple interfaces with different d-spacing regions in MEGNIs grown on NAPS-X glasses (*21*). These observations confirm the significant influence of the support on tailoring hot carrier dynamics.

Figure S28A illustrates the variations in the $I_{490}/I_{530}$ intensity ratio with increasing $SiO_2$ substitution. The peak intensity ratio of $I_{490}$ to $I_{530}$ ($I_{490}/I_{530}$) increases (Figure S28A) with higher $SiO_2$ concentrations in the NAPS-X glass, suggesting a reduction in the bleaching signal intensity due to the broadening of the LSPR band and reduced absorption efficiency at LSPR region. The highest intensity ratio of 2.6 was observed for $SiO_2$-Au glass, where a broad LSPR was attributed to the larger size and weakly embedment of the islands. The substitution of $SiO_2$ for $P_2O_5$ in NAPS-0 glass (NAPS-5 and NAPS-15) shows a similar intensity ratio, while substituting $SiO_2$ for $Al_2O_3$ in NAPS-20 glass (NAPS-25 and NAPS-30) increases this ratio. Interestingly, for NAPS-15-Au glass, the $I_{490}/I_{530}$ intensity ratio decreases as the pump wavelength increases from 350 nm to 400 nm and then to 450 nm (Figure S28B). Under 350 nm illumination, it is evident that the positive peak at 490 nm arises from ESA and the inter-band transition from the d-band to the conduction sp-band. The combination of ESA and inter-band transition generates a large number of electrons at higher energy levels, making MEGNIs embedded NAPS-X glasses suitable for photocatalysis and photodetector applications under UV light illumination.

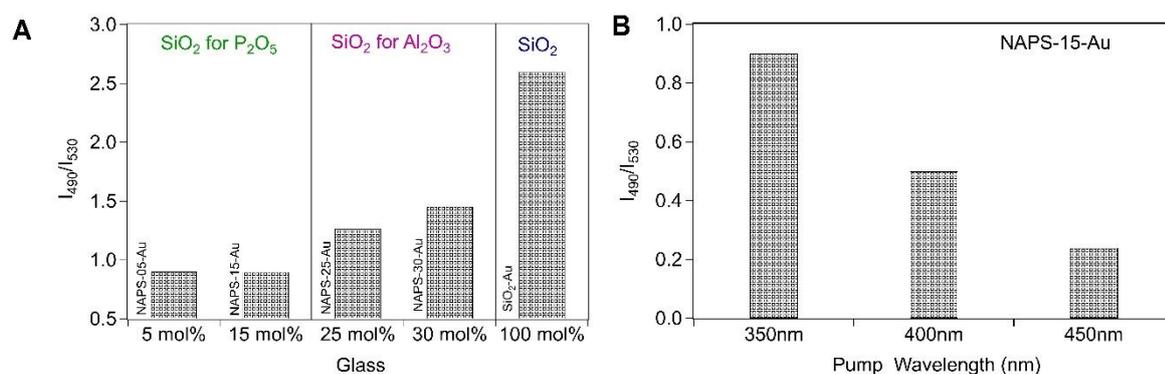

**Figure S28:** Peak intensity ratio of excited state absoprtion ($I_{490}$) and ground state bleaching ($I_{530}$) (A) with increasing the $SiO_2$ concentration in NAPS-0 glass at 350 nm pump wavelength and (B) with increasing the pump wavelength for NAPS-15-Au glass.

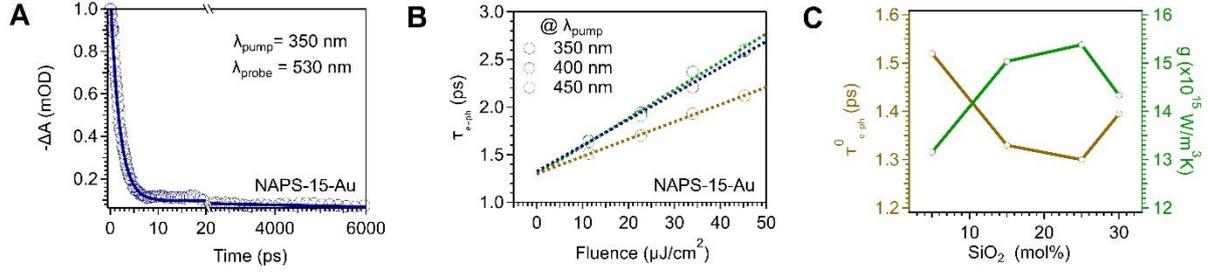

**Figure S29:** *(A) Kinetics of the normalised ground state bleaching signal at 530 nm. (B) Electron phonon coupling time for NAPS-15-Au glasses with increasing the pump fluence at different pump wavelengths. (C) electron phonon coupling time for NAPS-X-Au glasses at zero pump power and electron phonon coupling constant for NAPS-X-Au glasses.*

To quantitatively characterize energy dissipation pathways, we determined electron-phonon coupling constants (*g*) across all compositions. For each NAPS-X-Au sample, $\tau_{e\text{-}ph}$ exhibits linear dependence on pump power, validating the two-temperature model (*26*) and establishing the relationship:

$$\tau_{e\text{-}ph} = \gamma(T_0 + \Delta T_e)/g \qquad (S6)$$

where γ represents the electron heat capacity coefficient (66 J/m³K² for gold at room temperature $T_0$). By extrapolating to zero pump power ($\Delta T_e = 0$) (Figure 5D) and confirming wavelength independence of the zero-power electron-phonon coupling time $\tau^0_{e\text{-}ph}$ (Figures 29B and 29C), we calculated composition-specific g values ($g = \gamma T_0/\tau^0_{e\text{-}ph}$) for all samples (Figure S29C).

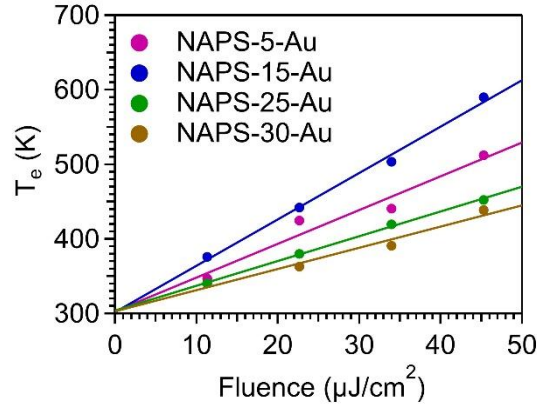

**Figure S30:** Electron temperature ($T_e$) variation with pump power for all NAPS-X-Au glasses. $T_e$ was calculated considering that electron phonon coupling constant is independent of the temperature based on the two temperature model.

In general, $\tau_{e\text{-}ph}$ of hot electrons guide the efficiency of transferring hot electrons to the surface-bound molecules and or to the integrated thin films. The obtained $\tau_{e\text{-}ph}$ are higher than 1.5 ps, indicating their suitability for the catalytic applications. Further, patterned gold nano-arrays have been shown to act as effective nucleating promoters for perovskite optoelectronic materials (*27*). The ultrastable MEGNIs on a transparent glass serve as nucleation centers, leading to the formation of perovskite with diverse morphologies. Additionally, it was observed that the peak intensity ratio of $I_{490}$ to $I_{530}$ ($I_{490}/I_{530}$) decreases (Figure S28) as the pump wavelength increases. This indicates that the MEGNIs on NAPS glasses are highly sensitive to the excitation wavelength and provide efficient hot carriers to the integrated semiconductors or perovskites in metal-semiconductor heterojunctions, especially upon UV exposure. Such properties pave the way for the design and development of commercially viable optoelectronic, sensing, and photocatalytic materials over a wide range of applications.